\documentclass[letterpaper,english,twocolumn, aps,nofootinbib,pre,showpacs]{revtex4-2}
\usepackage[OT2,T1]{fontenc}
\usepackage[latin9]{inputenc}
\usepackage{bm}
\usepackage{amsmath}
\usepackage{amssymb}
\usepackage{graphicx}
\usepackage{xcolor}

\makeatletter
\DeclareFontFamilySubstitution{OT2}{\rmdefault}{wncyr}
\DeclareRobustCommand{\rn}[1]{
	{\fontencoding{OT2}\selectfont#1}%
}

\pdfpageheight\paperheight
\pdfpagewidth\paperwidth

\makeatother

\usepackage{babel}
\begin{document}
\title{Phase diagram of generalized TASEP on an open chain: Liggett-like
boundary conditions}
\author{Nadezhda Zh Bunzarova}
\affiliation{Institute of Mechanics, Bulgarian Academy of Sciences, Sofia, Bulgaria }
\author{Nina C Pesheva}
\affiliation{Institute of Mechanics, Bulgarian Academy of Sciences, Sofia, Bulgaria }
\author{Alexander M Povolotsky}
\affiliation{Bogoliubov Laboratory of Theoretical Physics, Joint Institute for
Nuclear Research, Dubna, Russia }
\affiliation{National Research University Higher School of Economics, Moscow, Russia }

\begin{abstract}
The totally asymmetric simple exclusion process with generalized update is a version of the discrete time totally asymmetric exclusion process with an additional inter-particle   interaction that controls the degree of particle clustering. Though the model was shown to be integrable on the ring and on the infinite lattice,  on the open chain it was studied mainly numerically, while no analytic results existed even for its  phase diagram.  In this paper we introduce  new boundary conditions associated with the infinite translation invariant stationary states of the model, which allow us  to obtain  the exact phase diagram analytically. We discuss the phase diagram in detail and confirm the analytic predictions  by extensive numerical simulations.   
\end{abstract}

\maketitle

\section{Introduction}

The asymmetric simple exclusion process (ASEP) is the paradigmatic
model of the non-equilibrium statistical physics. It captures main
features of a variety of driven diffusive systems and has plenty of applications
to many phenomena from traffic flows to interface growth as well as
to transport in biological and chemical systems, \cite{halpin1995kinetic,krug1997origins,schadschneider2010stochastic}. Being a toy model,
it nevertheless plays a key role as a laboratory for studies of non-equilibrium
stationary states \cite{derrida2007non}, boundary induced phase transitions \cite{krug1991boundary}and Kardar-Parisi-Zhang
theory of universal fluctuations and correlations in driven stochastic
systems  \cite{kardar1986dynamic}.

ASEP is a system of particles on a one-dimensional lattice performing
stochastic jumps to nearest neighbor sites subject to exclusion interaction.
The model was first proposed in a biological context to model the
mRNA translation in protein synthesis in 1968 \citep{macdonald1968kinetics}.
Two years later it was introduced in mathematical literature as a
model of interacting Markov processes \citep{spitzer1970interaction}.
Since then ASEP was a subject of numerous studies. Detailed results
were obtained on its stationary state in finite periodic  and infinite
systems  as well as  on a finite segment with open boundaries connected
to particle reservoirs. Also, the non-stationary dynamics was analyzed
in the infinite system  and on the ring. Since the list of references is more than extensive, we will not give
a detailed account of it here. The main ideas can be learned from  reviews \cite{derrida1998exactly,golinelli2006asymmetric,lazarescu2015physicist,corwin2016kardar,remenik2022integrable} and the details can be found in the references therein.  The key reason why so many results could be obtained was a special integrable mathematical structure
of the model that makes such tools as Bethe ansatz and matrix product
ansatz (MPA) applicable to obtaining the exact solutions of the model \cite{schutz2001exactly}.

Originally the model was formulated in continuous time. Also several
discrete time generalizations of TASEP, the totally asymmetric version
of ASEP, were proposed. Being stochastic cellular automata they suit
well for computer simulations and were extensively used e.g. for modeling
traffic flows. Though continuous time models are simpler to analyze,
their discrete time descendants having richer dynamics allow testing
the universality of results obtained for continuous time models. Different
versions of discrete time TASEP use different update procedures, e.g.
random, backward and forward sequential, parallel and sub-lattice
parallel updates \cite{hinrichsen1996matrix,evans1999exact,de1999exact,brankov2000exact}. The stationary states on a ring and on a segment
were constructed for all these cases, see \cite{rajewsky1998asymmetric} for review and references therein. Qualitatively
the large scale pictures obtained from studies of continuous and discrete
time TASEPs are similar, despite the difference in their exact steady
states. How an inclusion of new types of interactions into the dynamics affects the typical
behavior of the system is an interesting problem and in particular
is a motivation for the present work.

In this paper we focus on the stationary state of a generalization
of discrete time TASEPs, TASEP with generalized update (gTASEP), considering
it on a segment with open boundary conditions (BC). The difference
of gTASEP from the usual discrete time TASEPs is the presence of an
extra interaction controlling the particle clustering. When the parameter
responsible for the interaction (probability of catching up the particle
ahead) is varying from zero to one, the model crosses over from TASEP
with parallel update through TASEP with backward sequential update
to the deterministic aggregation regime in which particles irreversibly
aggregate to giant clusters, each moving as a single particle.

{ 
The  addition of  extra interaction into TASEP dynamics can be useful for applications as it provides more flexibility in  the modeling real life phenomena, while retaining  the advantage of the model being  exactly solvable.  In particular, while  TASEP is commonly used as  one of basic models for traffic flow in a single-lane roadway \cite{schreckenberg1995discrete,schadschneider2010stochastic},  gTASEP 
 allows new aspects of drivers' behavior to be taken into account.
Similar issues are widely studied in the framework of so-called ``car-following'' models of traffic (for recent reviews see e.g. \cite{ahmed2021review,zhang2023review}), which include   a set of  parameters like maintaining appropriate gap, speed adoption, desired acceleration or deceleration, etc.,  used  for more realistic description of real traffic systems.  Similarly, the additional interaction in gTASEP allows a  study of traffic flow in different regimes, from a ``repulsive'',   when a car is lagging behind the car ahead, to an ``attractive'',  when the drivers  tend to  catch up with the car ahead trying    to maintain the same speed as that car.  Thus, clusters of synchronously moving cars may appear, leading to higher throughput. In the limiting case of irreversible aggregation  the clusters of simultaneously moving cars, move forward as a whole entity. This is a simplified picture of synchronously moving vehicles in a jam.
	
Also, gTASEP is  one of the simplest models, which allows a study of the aggregation-fragmentation processes in one dimension \cite{brankov2018model,bunzarova2017one}. Depending on the  nature of particles, they may attract or repel each other (if e.g. they are charged), or they may be neutral. The examples of physical systems in nature, in which competing aggregation and  fragmentation   subject to   diffusion  play a major role, like e.g. reversible polymerization in solutions  \cite{blatz1945note} and coagulation of colloidal particles \cite{family2012kinetics},  are numerous  appearing in chemistry, biology and physics. 
}

gTASEP was first proposed in \citep{woelki2005steady} as a generalized
version of TASEP that can be mapped to a zero-range type model with
factorized steady state. This mapping allowed a study of translation
invariant stationary states on a ring and its limit to the infinite
system. Later, it was rediscovered as a Bethe ansatz solvable two-parameter
generalization of TASEP \citep{derbyshev2012totally} and as a particular
limit of more general integrable three parametric chipping model \citep{povolotsky2013integrability}.
It also appeared in the studies of the Schur measures related to deformations
of the Robinson-Schensted-Knuth dynamics \citep{knizel2019generalizations}.
The integrable structure of the process allowed obtaining exact solutions
for the stationary state and for the current large deviations on the
finite periodic lattice \citep{derbyshev2015emergence} as well as
 for the non-stationary dynamics on the infinite lattice \citep{derbyshev2021nonstationary,knizel2019generalizations}.
In particular, these studies revealed the crossover from the Kardar-Parisi-Zhang
universality class to the deterministic aggregation regime that
takes place on the diffusive scale.

Yet, gTASEP in a finite system with open BC has still resisted an
exact analytic treatment. Note that the stationary state of driven
diffusive systems connected to particle reservoirs is usually highly
nontrivial. The first exact solution for the stationary state of TASEP
on an open chain was found using recursion relations between the steady
states in systems of different sizes \citep{derrida1992exact}. This
solution inspired a discovery of the matrix product method \citep{derrida1993exact}
that received tremendously wide range of applications and became a
basic tool for exact construction of stationary states of non-equilibrium
lattice gases \citep{blythe2007nonequilibrium}. In particular, following
the solution of the continuous time model \citep{derrida1993exact,schutz1993phase}
the stationary states of discrete time TASEPs with parallel and ordered
sequential updates were obtained \citep{rajewsky1998asymmetric,de1999exact,brankov2000exact}.
Using these solutions exact phase diagrams were constructed.

No any such results are available for gTASEP. An attempt \citep{aneva2016matrix}
of finding the matrix product steady state for it led only to the
trivial two-dimensional representation of the bulk matrix algebra
that produced the known Ising-like stationary measure on the ring
\citep{derbyshev2015emergence}. On the open chain extensive numerical
simulations supported with random walk based heuristic arguments were
used to study the deterministic aggregation limit of the model \citep{bunzarova2017one,brankov2018model}
as well as the whole phase diagram \citep{bunzarova2021aggregation},
for which, however, no any conclusive analytic predictions were still
made. In this paper we make the first step in this direction obtaining
conjecturally exact formulas for the phase diagram of gTASEP on a
finite segment with the so called Liggett-like BC introduced below.

{ Introduction of these new BC to gTASEP is the key idea of the present paper. 
Roughly, they are defined in such a way that the    particle injection to and ejection from the system  at the ends of the finite lattice have the same probabilities as similar   particle jumps in the infinite  translation invariant stationary states of gTASEP.  The structure of such  stationary states parameterized by the density of particles   is  known and can be analyzed with usual statistical physics toolbox.  In this way we can identify the so-called  high and low density phases, governed  by the   particle injection and ejection respectively, and   construct the phase diagram of the model  using purely hydrodynamic arguments. }

The paper is organized as follows. In section \ref{sec: Cluster dynamics} we 
define the model including the  new rule of the update of boundary site, 
which is the key formula of the present paper. {In section  \ref{sec: Phase diagram of TASEP} we describe a typical phase diagram of usual TASEP  previously obtained from exact solutions, explain how it can be obtained for TASEP with Liggett-like BC from the known structure of the stationary state using only hydrodynamic arguments and outline the further steps of the same program applied to gTASEP.} In section  \ref{sec: Boundary param} we describe the stationary 
state of gTASEP on the ring of finite size,  which  is then assumed to  grow to infinity, and calculate 
probabilities of events that should mimic  the particle injection  to and ejection from 
the finite system.  Using the formulas obtained we construct the phase diagram in section \ref{sec: Phase diagram}. 
The results of numerical simulations and their comparison with those for gTASEP with the previously studied version of   boundary update rules  are presented in section  \ref{sec: Simulation results}. We make a few concluding comments and discuss the perspectives in section \ref{sec: Conclusion and discussion}.

\section{Definition of the model \label{sec: Cluster dynamics}}

Let us consider a 1D lattice $\mathcal{L}=\{1,\dots,L\}$ with $L$
sites or the infinite lattice $\mathcal{L}=\mathbb{Z}$. A configuration
$\boldsymbol{\tau}=\{\tau_{i}\}_{i\in\mathcal{L}}\in\{0,1\}^{\mathcal{L}}$
of gTASEP is a binary string of occupation numbers, which take the
values $\tau_{i}=1$ if site $i$ is occupied by a particle and $\tau_{i}=0$
if it is empty. gTASEP evolves in discrete time. At every time step
the whole particle configuration is updated using so called cluster-wise
sequential update. By particle cluster we mean a sub-string $(\dots01^{n}0\dots)$
of $n\geq1$ consecutive sites filled with particles surrounded by
two empty sites on both sides. The clusters of this form are referred
to as bulk clusters. If we consider the finite lattice with open boundary
conditions, also boundary clusters of the forms $(1^{n}0\dots)$ and
$(\dots01^{n})$ with $n\geq1$ may exist. In them the empty site
that would fall beyond left or right boundary respectively is omitted.
The cluster that touches both boundaries completely fills the whole
lattice being left and right boundary cluster simultaneously. Then, no bounding empty sites are assumed. 

{\subsection{Bulk dynamics}}

\begin{figure}

\includegraphics[width=0.7\columnwidth]{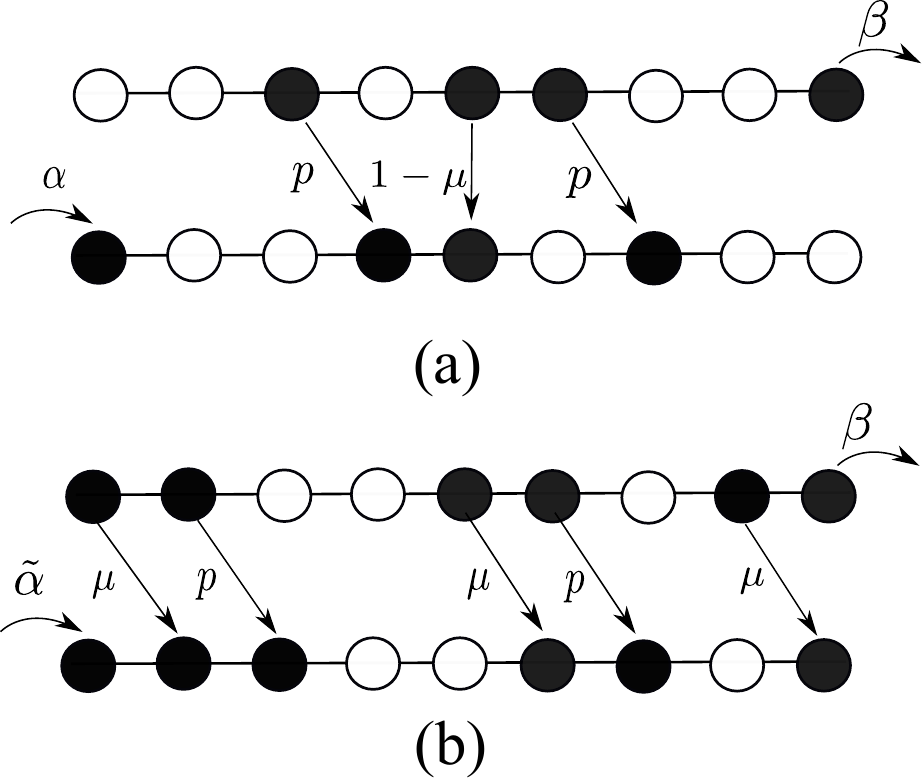}\caption{Two instances of  gTASEP configuration updates  having probabilities (a) $\alpha p^2(1-\mu)\beta$ and (b) $\tilde{\alpha} p^2 \mu^3 \beta$. A particle jumps from a lattice site to the next site on the right with probability $p$ or stays  with probability $(1-p)$ if the target site was empty before the update. The probabilities change to  $\mu$ and $(1-\mu)$ respectively if the target site  has emptied at the same step. Similarly in  (a) a particle enters to the leftmost site that was empty before the update with probability $\alpha$ and in (b) the probability is $\tilde{\alpha}$, since  the site was occupied. The particle in the rightmost site leaves the lattice  with probability  $\beta$.	
\label{fig: gtASEP}}

\end{figure}

The cluster-wise sequential update procedure is as follows, see Fig.
\ref{fig: gtASEP}.
 At every time step, all clusters are updated simultaneously and independently
according to the following rules. From a bulk cluster the rightmost
particle jumps to the next site on the right with probability $0< p\leq1$.
If the rightmost particle has jumped, the next one jumps with probability
$0\leq\mu\leq1$, and so does every next particle within the cluster
provided that the previous particle has jumped. The update of the
cluster ends either when one of its particles has decided not to jump
or all particles of the cluster has moved one step to the right.

In other words, given a cluster $(\dots01^{n}0\dots)$ with $n$ particles,
its $0\leq m\leq n$ rightmost particles move a step to the right,
\[
(\dots01^{n}0\dots)\to(\dots01^{n-m}01^{m}\dots),
\]
with probability
\begin{equation}
\varphi(m|n)=\frac{v(m)w(n-m)}{f(n)},\label{eq: phi(m|n) - general}
\end{equation}
 defined in terms of two functions
\begin{eqnarray}
v(k) & = & \mu^{k}(\delta_{k,0}+(1-\delta_{k,0})(1-\nu/\mu)),\label{eq:v(k)}\\
w(k) & = & \delta_{k,0}+(1-\delta_{k,0})(1-\mu),\label{eq:w(k)}
\end{eqnarray}
{were, we have introduced yet another parameter $\nu$ related to $p$
and $\mu$ by identity
\[
p=\frac{\mu-\nu}{1-\nu}.
\]
}The denominator 
\begin{equation}
f\left(n\right)=\sum_{i=0}^{n}v(i)w(n-i),\label{eq: f(n) form}
\end{equation}
is the normalization factor given by
\begin{equation}
\label{eq: f(n)}
f\left(n\right)=\delta_{n,0}+(1-\delta_{n,0})(1-\nu).
\end{equation}
For $p,\mu$ to be probabilities, $0\leq p,\mu\leq1$, parameter $\nu$
should take values in the range $\text{\ensuremath{\nu\leq\mu}}.$ {Substituting $m=0,1,\dots,n$  into (\ref{eq:v(k)}-\ref{eq: f(n)})  one can directly check that these formulas are consistent with the dynamical rules described in the beginning of the section.

 The form (\ref{eq: phi(m|n) - general}) of the hopping probabilities comes from  \cite{evans2004factorized}, 
where it was shown to be a necessary and sufficient condition of the factorized form of the
stationary probability in the so-called chipping model with zero range interaction on the ring. In that model  an arbitrary number of particles in a site is allowed  and $\varphi(m|n)$ is a probability for $m$ particles to jump out of a site with $n$ particles to the next site on the right. 
Models of this type (ZRP-like) can be related  to TASEP-like models, in which a site can
be occupied by at most one particle, by the so-called ZRP-ASEP
mapping that replaces a cluster of particles with a single site
with the same number of particles, while the one-step move of a part
of the cluster to the right corresponds to the jump of as many particles
to the next site. This mapping was first  used to study the stationary state of the discrete time TASEP \cite{evans1999exact} 
and gTASEP \cite{woelki2005steady} on the ring.

In  formulas (\ref{eq: phi(m|n) - general}-\ref{eq: f(n)}) specifying   probabilities $\varphi(m|n)$ of gTASEP one can recognize a particular
$q=0$ limit of jumping probabilities of the integrable chipping model
introduced in \citep{povolotsky2013integrability} in the search for the Bethe ansatz solvable version of zero range chipping model from \cite{evans2004factorized}, also studied in
\citep{borodin2015spectral} under the name $q-$Hahn TASEP. The notation
$p_{m}$ for the probability $\mu$ of catching up the particle ahead
was also used in \citep{bunzarova2017model,bunzarova2017one,brankov2018model,bunzarova2021aggregation}.
Here we keep to $p,\mu,\nu$ notations of \citep{povolotsky2013integrability,derbyshev2015emergence}.
}

The bulk rules described specify completely the update procedure of
gTASEP on the finite lattice with periodic boundary conditions, $\tau_{i}\equiv\tau_{i+L}$,
and on the infinite lattice. In particular, $\mu=0$ and $\mu=p$
limits reproduce the rules of usual TASEP with parallel update (PU) and backward sequential update (BSU) respectively.
In both  cases all particles jump with the same probability $p$, with  the sites being updated simultaneously in the former case and sequentially from right to left in the latter.  We also refer to the cases $\mu<p$ and $\mu>p$ as  repulsion and attraction regimes, since the particle clustering in the stationary infinite system  decreases and increases respectively comparing to the independent Bernoulli stationary state with the same density  of the $p=\mu$ BSU case (See \cite{derbyshev2015emergence} and discussion in the following sections).   

The $\mu=1$ limit is the deterministic aggregation regime, in which
particles that once got into the same cluster never split up, while
the clusters themselves move as isolated particles and, when they meet,
merge into bigger clusters.

Note that the cluster-wise update defined coincides with the familiar
BSU prescription  in many cases, e.g. in the infinite system with
particle configurations bounded from the right as well as in the finite
chain with open BC considered below. However, we prefer to use this
formulation, because of subtleties which appear e.g. on the ring,
where an ambiguity in the choice of the starting point of update exists.
{Also, we recall that  TASEP-like models with such an update  are mapped under the ZRP-ASEP mapping to   the  chipping models with zero-range interaction and   parallel update from  \citep{evans2004factorized}, which  become tractable, when the hopping probabilities have the form (\ref{eq: phi(m|n) - general})}.

{\subsection{ Boundary dynamics}}

If gTASEP is considered on the finite segment, the updates of the
boundary clusters should be defined separately. We will assume that
particles may be added into the system at the leftmost site, $i=1$,
of the lattice and removed from the system from the last site, $i=L$.
On the update of the boundary cluster near the right boundary the
particle at site $i=L$ is removed with probability $\beta$. If this
has happened and the cluster consists of more than one particle, the
next particle makes a step to the emptied site with probability $\mu$
and so does the next particle, etc, until either all particles of
the cluster have made a step or some particle decides not to jump
with probability $(1-\mu)$. Thus, the possible outcomes of an update
of the right boundary cluster with $n\geq1$ particles have the following
probabilities
\begin{align*}
(\dots01^{n})\to(\dots01^{n}): & \,1-\beta,n\geq1;\\
(\dots01^{n})\!\to\!(\dots01^{n-m}01^{m-1}): & \,\beta\mu^{m-1}(1-\mu),1\leq m<n;\\
(\dots01^{n})\to(\dots001^{n-1}): & \,\beta\mu^{n-1}.
\end{align*}

How is the left boundary updated? If the leftmost site, $i=1$, of
the lattice was empty before the update, a particle is added to this
site with probability $0<\alpha\leq1$, so that the left boundary
cluster appears. If the left boundary cluster existed, it is updated
in the same way as the bulk cluster. If the leftmost site $i=1$ has
emptied as a result of this update, a particle may be added there
with yet another probability

\begin{equation}
\tilde{\alpha}=\frac{\alpha\mu}{\mu-(1-\alpha)\nu}=\frac{\alpha\mu(1-p)}{\alpha(\mu-p)+p(1-\mu)}.\label{eq:tilde alpha-1}
\end{equation}
Then the updates of the left boundary with corresponding probabilities
are
\begin{align*}
(0\dots)\to(1\dots) & :\,\alpha;\\
(0\dots)\to(0\dots) & :\,1-\alpha;\\
(1^{n}0\dots)\to(1^{n}0\dots) & :\,1-p,1\leq n;\\
(1^{n}0\dots)\to(1^{n-m}01^{m}\dots) & :\,p\mu^{m-1}(1-\mu),1\leq m<n;\\
(1^{n}0\dots)\to(01^{n}\dots) & :\,p\mu^{m-1}(1-\tilde{\alpha}),1\leq n;\\
(1^{n}0\dots)\to(1^{n+1}\dots) & :\,p\mu^{n-1}\tilde{\alpha},m<n.
\end{align*}
If the  cluster to be updated fills the whole lattice, the probabilities are prescribed  in the same way as to the 
update of the left boundary cluster up to the change of the first particle jump probability  $p$  to $\beta$.

{Relation    (\ref{eq:tilde alpha-1}) between  $\alpha, \tilde{\alpha}$ and the bulk jump probabilities $p,\mu$ is the
key formula of the present paper. Given here as a definition, 
it will be derived below from the assumption that  the injection probabilities coincide with the probabilities  of similar particle jumps in 
the infinite stationary translation invariant system.}
One can see that given fixed values
of $p$ and $\mu$, parameter $\tilde{\alpha}$ is the monotonous
function of $\alpha$ taking values in the range $$0\leq\tilde{\alpha}\leq1$$
as $\alpha$ varies in the same range. Furthermore, as will be explained
below, when the values of $\alpha$ vary in the range
\[
0<\alpha\leq p,
\]
and consequently, $\tilde{\alpha}$ following (\ref{eq:tilde alpha-1})
takes values in $$0\leq\tilde{\alpha}\leq\mu,$$ 
these parameters can be associated with the stationary state of the infinite  translation invariant
system with the same bulk dynamics at some particle density value. The same applies to the right
boundary condition with parameter $\beta$ in the rage $0\leq\beta\leq p$.
Thus, we refer to so defined boundary dynamics as  Liggett-like
BC. All the results obtained  below  for  gTASEP on a segment  are the consequence of this definition.
 
In particular cases of PU and BSU bulk dynamics, corresponding to
$\mu=0$ and $\mu=p$ respectively, we find $\tilde{\alpha}=0$ and
$\tilde{\alpha}=\alpha$, the values that reproduce the corresponding
BC that were used in the previous exact solutions of those models.
We would expect that the choice (\ref{eq:tilde alpha-1}) would also
be in some sense good, i.e. potentially exactly solvable. 

\begin{figure}
	\includegraphics[width=0.48\columnwidth]{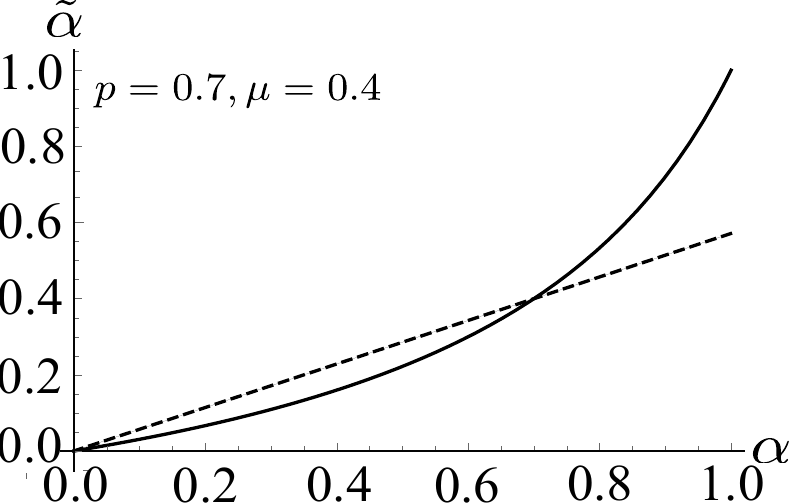}\,\,
	\includegraphics[width=0.48\columnwidth]{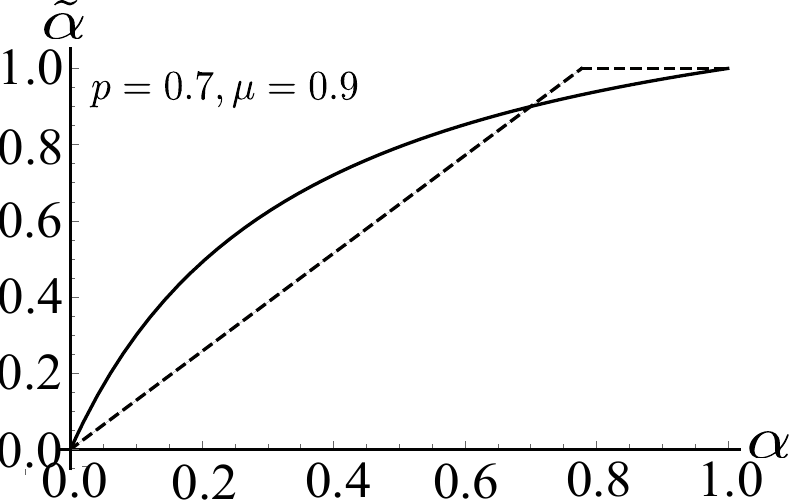}\\
	(a) 	\hspace{0.4\columnwidth} (b)
	\caption{Typical  plots of $\tilde{\alpha}$  vs $\alpha$ in (a) attracting, $p<\mu$, and (b) repulsive, $p>\mu$, regimes for two  definitions of $\tilde{\alpha}$:   formula (\ref{eq:tilde alpha-1}) -- solid curves,  and formula  (\ref{eq: Standard BC}) -- dashed curves.\label{eq: tilde alpha vs alpha}}
\end{figure}

The formula
(\ref{eq:tilde alpha-1}) is to be compared with another choice of
formula
\begin{equation}
\tilde{\alpha}=\min(1,\alpha\mu/p)\label{eq: Standard BC}
\end{equation}
proposed in \citep{phdthesis,bunzarova2017one}. This prescription
also matches with the of PU and BSU cases. However, in general it
is not clear how to associate such BC with the fixed density reservoir,
which makes the analysis of phase diagram a more difficult problem.

To compare them qualitatively, we note  that at fixed $p>\mu$  ($p<\mu$)  $\tilde{\alpha}$ defined by  (\ref{eq:tilde alpha-1}) is a monotonously increasing  convex (concave) function of $\alpha$.  Its plot has a unique common point  with the piece-wise linear function (\ref{eq: Standard BC}) in the interval $\alpha\in (0,1)$ at $\alpha=p$.  {Thus,   in the repulsive regime, $p>\mu$,  the effective flow into the system   with BC (\ref{eq:tilde alpha-1})  is enhanced  compared to the that  with BC (\ref{eq: Standard BC})  when  $\alpha<p$ and suppressed  when $\alpha>p$  and vice versa in the attractive regime,   $p<\mu$, see fig. \ref{eq: tilde alpha vs alpha}. } Below will see  numerical consequences of this fact.\\

{\section{Phase diagram of TASEP \label{sec: Phase diagram of TASEP}}

Before describing our results let us informally discuss  how the known 
phase diagrams obtained from exact solutions of the usual TASEP typically look like and what is the significance of the Liggett-like BC.  The examples, which we keep in mind, are the discrete time TASEPs with  BSU  and PU, obtained from the  above defined gTASEP by setting  $\mu=p$ and $\mu=0$ respectively, and  the continuous time TASEP that can be obtained from both discrete time  cases in $p\to 0$ limit.

In the   usual continuous (discrete) time TASEPs all particles 
on a 1D lattice jump in one direction with 
same unit rate (probability $p$) subject to the exclusion
interaction.  Also  boundary rates (probabilities), $\alpha,\beta$,
which are in general different from the bulk one, are assigned to
particle injection to the first site of the chain and particle ejection
from the last site of the chain, which also happen according to the same update.  
This is the simplest and most natural  choice of BC. 

By luck, in a suitable  range of $\alpha,\beta$ so  introduced BC, are the examples of the BC
introduced by Liggett \citep{liggett1975ergodic,liggett1977ergodic}.
They imply that the leftmost (rightmost) site of the segment is attached
to a reservoir, which is the left (right) half of a 1D infinite system with the same TASEP
being in the translation invariant steady state with fixed particle
density value related to the value of boundary parameter $\alpha$
($\beta$). (Note that every mentioned version of TASEP on the infinite
chain has a family of rather simple translation invariant steady states
parameterized by the value of particle density. These are product
of independent Bernoulli measures in continuous time as well as the BSU
case and 1D Ising-like measure in the PU case.) Obviously, when the
densities of the right and left reservoirs are equal, the segment
looks simply as a part of the infinite system with TASEP in the translation
invariant steady state with fixed particle density value.
When the densities of boundary reservoirs are different, the system maintains the least of  the  currents    the two stationary states of the reservoirs support.  The  state of the winning  reservoir then propagates through almost all the system, which is then either in  high density (HD) or  low density (LD) phase corresponding to  the left or right reservoir respectively.  Also,  the values of both $\alpha$ and $\beta$ can be chosen  greater than those associated with  any stationary state of the infinite system. Then,  the system is in the maximal current (MC) phase maintaining  the maximal value of  current the infinite stationary system is able to support.

Thus, the typical plot of a simplest phase diagram in $\alpha-\beta$
plane observed before  looks as shown in fig. \ref{fig: phase diagram TASEP}.
It consists of three domains, LD, HD and MC phases, in which the stationary density of
particles is either dominated by the left reservoir or by the right
reservoir or maximizes the bulk current respectively. The LD-MC and
HD-MC phase transitions are continuous (in density) and HD-LD phase
transition is the first order phase transition.
\begin{figure}
	\includegraphics[width=0.9\columnwidth]{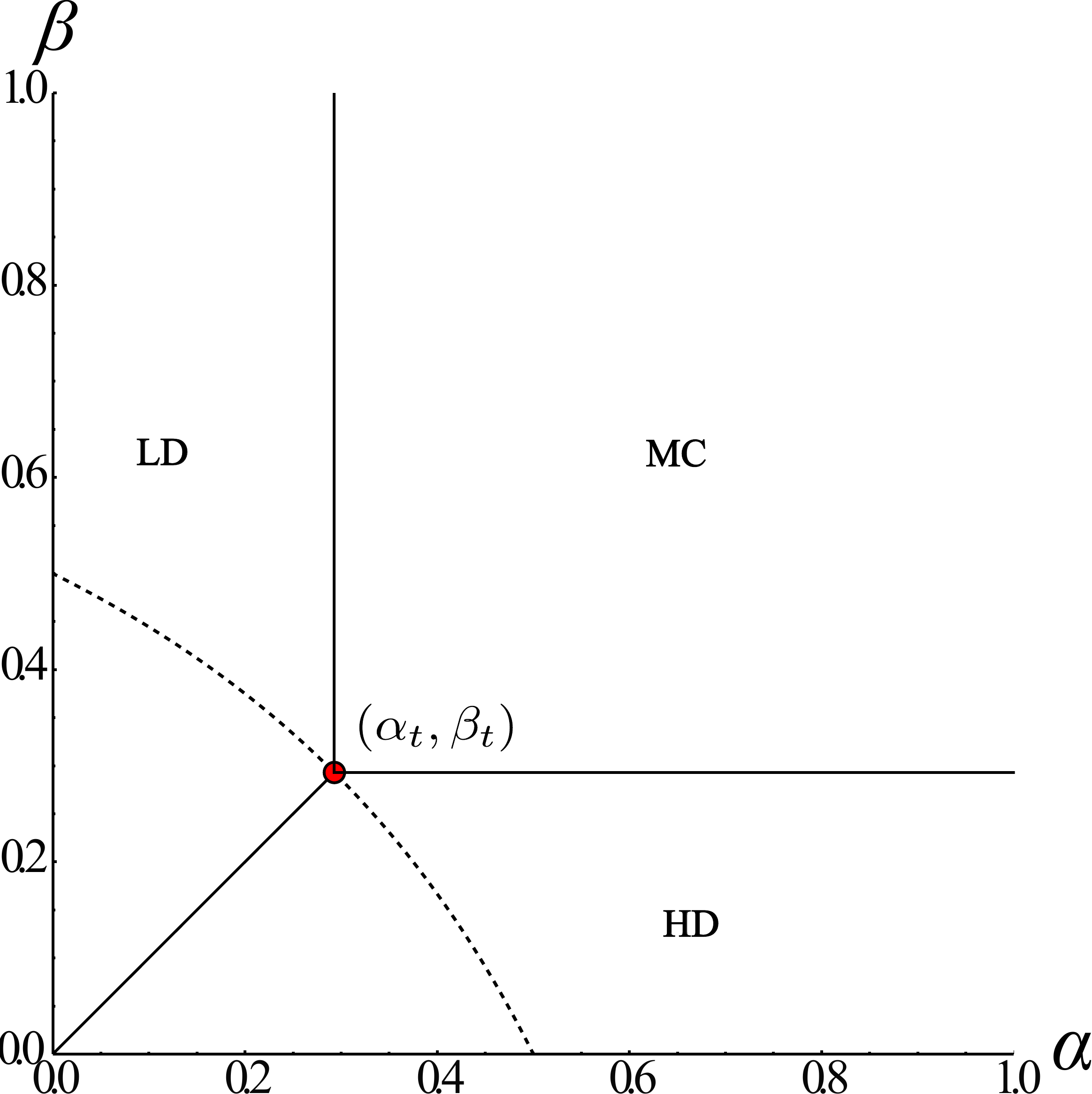}\caption{Typical phase diagram of TASEP-like system with three phases: HD,LD
		and MC. The HD-LD phase coexistence line is straight for known exactly-solvable
		TASEPs. The stationary state trivializes on the dashed line. \label{fig: phase diagram TASEP}}
\end{figure}
The phase coexistence lines are straight lines linking the points
$(0,0)$, $(1,\beta_{t})$ and $(\alpha_{t},1)$ to the triple point
$(\alpha_{t},\beta_{t})$. The values of $(\alpha_{t},\beta_{t})$
are equal to $\alpha_{t}=\beta_{t}=1/2$ in continuous time case and
depend on the value of $p$ and on the update in discrete time. Also,
there is a special curve in the phase diagram, corresponding to the
mentioned regime with equal densities of the right and left particle
reservoirs, where the stationary state trivializes, i.e. the finite
segment looks as the part of translation invariant stationary infinite
system. The points of this curve belong to either LD or HD phases, with
the only exception, the point where the density attains its maximum
equal to the density of MC phase, which is the triple point.

It is  not difficult to draw  the phase diagram described, given that the structure of the fixed  density translation invariant stationary 
states of the infinite TASEP is known.  Due to the choice of the Liggett-like BC the stationary state of the large system 
in LD (HD) phase looks  like that of the left (right) reservoir at almost all the lattice 
except for the close vicinity of the right (left) end. It is simply the stationary state 
current density relation that allows one to associate every pair of $(\alpha,\beta)$
with a certain macroscopic phase at least in the case of simplest three phase scenario. This
scenario is known to be realized, when the infinite system current-density
relation is convex, which, as will be seen below, is also the case for gTASEP.

Note, that imposing BC in gTASEP on a finite segment is a more delicate
problem than in particular cases we just described because of the additional parameter $\mu$, probability of catching
up the particle ahead, controlling particle clustering in the bulk.
The way how one can incorporate an analogue of this parameter into
the boundary dynamics is not unique. Probably the simplest candidate, Eq. (\ref{eq: Standard BC}),
was proposed in \citep{phdthesis,bunzarova2017one} in an ad hoc manner
consistent with the PU and BSU limits. However, no analytic predictions
were made for those BC probably because it was not clear how to associate
such BC with  fixed density reservoirs. What allows us to partly fill this gap is the introduction of the Liggett-like boundary conditions for gTASEP.

Below we  complete the following program.
We first derive the  Liggett-like boundary conditions,    Eq. (\ref{eq:tilde alpha-1}). 
To this end, we prepare the infinite stationary gTASEP starting from the system on a finite ring and calculate the probabilities   of events corresponding to particle injection  to  and ejection from the system that is a finite part of the infinite stationary system  at a given particle density. In this way, the ejection and injection probabilities are associated with the particle density values in the boundary reservoirs. In course of this derivation we also discuss  calculation of averages of  observables  like correlation functions over the stationary state. In particular, we  reproduce the known current density relation, which is then used to construct the full phase diagram as described above.    

Assuming that the corresponding reservoirs
govern the stationary states in LD and HD phases we obtain the explicit
formulas of the triple point and of the coexistence line, which is
enough to describe the full phase diagram. Remarkably, unlike the
usual TASEPs studied before, the LD-HD phase coexistence line in gTASEP with Liggett-like BC
is not a straight line anymore. For generic bulk dynamics it is a
smooth curve connecting the origin and the triple point becoming the
straight line exactly in two particular limits of TASEP with BSU and
with PU. Finally, we will discuss how the phase diagram degenerates in the deterministic
aggregation limit. We also present  extensive numerical
simulations  which confirm the analytic results obtained.}

\section{Boundary parameters and stationary states of the infinite system \label{sec: Boundary param}}

As we explained above, thermodynamic properties of LD
and HD phases of gTASEP on a finite segment can be characterized in
terms of those of the stationary state of the infinite system maintaining
the same density as the density associated to the particle reservoirs
attached to the ends of the segment. To this end, we first remind
the construction of the stationary state developed in \citep{derbyshev2015emergence}
and then calculate the probabilities of events, corresponding to injection
and ejection probabilities to and from the boundary sites of the system.
The phase diagram is constructed by juxtaposition of the current-density
dependence and these probabilities. 

{ \subsection{Stationary state of gTASEP: from a finite ring to the infinite system.}}

To prepare the infinite system we start from the system on a finite
ring. The stationary measure was initially obtained from the mapping
of gTASEP to the system with zero range interaction that admits an
arbitrary number of particles in a site. At every time step $m$ particles
jump to the next site from a site with $n\geq m$ with probability
of the form (\ref{eq: phi(m|n) - general}), all sites being updated
simultaneously. The unnormalized stationary weight of a site with
$n$ particles in such a system is then given by the jumping probability
normalization $f(n)$ from (\ref{eq: f(n) form}) \citep{evans2004factorized}.
To translate this to the TASEP language, we should replace an $n$-particle
site with $n$ sites occupied by one particle each plus one empty
site ahead. Though this ZRP-ASEP mapping between configurations of
two models is not one to one, rather the correspondence is modulo
translations, the weight $f(n)$ is then assigned to $n$-particle
clusters, while the empty sites bring the unit weight.

Thus, the probability of particle configurations in gTASEP on the
ring with $L$ sites and $M<L$ particles depend only on their cluster
structure. Specifically, the probability of configuration
\begin{equation}
\bm{\mathbf{\tau}}=\left(0^{k_{1}}1^{n_{1}}\dots0^{k_{N_{c}(\bm{\mathbf{\tau}})}}1^{n_{N_{c}(\bm{\mathbf{\tau}})}}\right)\label{eq:tau}
\end{equation}
with $N_{c}(\bm{\mathbf{\tau}})$ clusters of sizes $n_{1},\dots,n_{N_{c}(\bm{\mathbf{\tau}})}\geq1$
interlaced with gaps of the lengths $k_{1},\dots,k_{N_{c}(\bm{\mathbf{\tau}})}\geq1$
is given by
\begin{align}
P_{_{st}}(\bm{\mathbf{\tau}}) & =\frac{\delta_{k_{1}+\cdots+k_{N_{c}(\bm{\mathbf{\tau}})},L-M}\delta_{n_{1}+\dots+n_{N_{c}(\bm{\mathbf{\tau}})},M}}{\mathcal{Z}(M,L)}\prod_{l=1}^{N_{c}(\bm{\mathbf{\tau}})}f(n_{l}),\label{eq:Can}\\
 & =\frac{\left(1-\nu\right)^{N_{c}(\bm{\mathbf{\tau}})}}{\mathcal{Z}(M,L)}\delta_{k_{1}+\cdots+k_{N_{c}(\bm{\mathbf{\tau}})},L-M}\delta_{n_{1}+\dots+n_{N_{c}(\bm{\mathbf{\tau}})},M}\nonumber
\end{align}
where the one-cluster factor is $f(n)=(1-\nu)$ for $n\geq1$ and
$\mathcal{Z}(M,L)$ is the canonical partition function, i.e. normalization
constant that ensures the unit sum of probabilities.

In the thermodynamic limit
\begin{equation}
M,L\to\infty,M/L=c,\label{eq:thermodynamic limit}
\end{equation}
the stationary distribution of local events depending on finitely
many sites coincides with the grand-canonical distribution that assigns
 a fugacity $z\geq0$ to a particle and an extra weight $f(n)=(1-\nu)$
to every cluster, so that the resulting weight of an $n$-particle
cluster is
\[
w_{z}(n)=z^{n}(1-\nu),
\]
while an empty site is assigned the unit weight. To calculate probabilities
of local events in the infinite system we start from the finite periodic
system with $L$ sites, where the number of particles is not fixed,
while the probability of a particle configuration (\ref{eq:tau})
with the number of particles $M(\bm{\tau})=n_{1}+\dots+n_{N_{c}(\bm{\mathbf{\tau}})}$
is
\begin{equation}
P_{z}(\bm{\tau})=\frac{1}{\mathcal{Z}_{L}(z)}\prod_{l=1}^{N_{c}(\bm{\mathbf{\tau}})}w_{z}(n_{l})=\frac{1}{\mathcal{Z}_{L}(z)}(1-\nu)^{N_{c}(\bm{\mathbf{\tau}})}z^{M(\bm{\tau})},\label{eq:GC}
\end{equation}
where $\mathcal{Z}_{L}(z)$ is the partition function. Of course,
in the finite system the grand-canonical distribution (\ref{eq: alpha-beta line})
is different from the exact canonical stationary distribution (\ref{eq:Can})
of gTASEP. They become equivalent only in the thermodynamic limit
(\ref{eq:thermodynamic limit}). The grand-canonical probability can
be evaluated using the transfer matrix formalism. Specifically, the
probability of a particle configuration $\bm{\tau}=(\tau_{1},\dots,\tau_{L})$
with occupation numbers $\tau_{1},\dots,\tau_{L}=0,1$, is
\[
P_{z}(\bm{\tau})=\frac{1}{\mathcal{Z}_{L}(z)}T_{\tau_{1},\tau_{2}}\dots T_{\tau_{L-1},\tau_{L}}T_{\tau_{L},\tau_{1}},
\]
where $T_{0,0}=1$, $T_{0,1}=T_{1,0}=\sqrt{z(1-\nu)},$ and $T_{1,1}=z$.
This measure is similar to the Gibbs measure of the 1D Ising model,
as was first observed in \citep{schreckenberg1995discrete} in the
context of the TASEP with PU.

Correspondingly, for the periodic BC the partition function is given
by the trace of $L$-th power of the transfer matrix
\[
\mathcal{Z}_{L}(z)=\mbox{Tr}T^{L}=\lambda_{1}^{L}+\lambda_{2}^{L},
\]
where $\lambda_{1},\lambda_{2}$ are the eigenvalues of the matrix
$T$ defined so that $\lambda_{1}>\lambda_{2}\geq0$

To proceed with the calculations it is convenient to introduce another
variable $z_{c}$ instead of $z$, which is a unique root of equation
\[
z=\frac{z_{c}\left(1-\nu z_{c}\right)}{1-z_{c}}
\]
 in the range
\begin{equation}
z_{c}\in[0,1].\label{eq:z_c range}
\end{equation}
The parameter $z_{c}$ has the meaning of particle fugacity in the
associated ZRP-like model and the further equations simplify significantly
written in terms of it, consisting of rational functions of $z_{c}$.
In particular the eigenvalues read as follows
\begin{eqnarray*}
\lambda_{1} & = & \frac{1-\nu z_{c}}{1-z_{c}},\quad\lambda_{2}=\nu z_{c}.
\end{eqnarray*}
 The largest eigenvalue $\lambda_{1}$ defines the specific free energy
in the thermodynamic limit (\ref{eq:thermodynamic limit})
\[
f(z)=\lim_{L\to\infty}\frac{\ln\mathcal{Z}_{L}(z)}{L}=\ln\lambda_{1}.
\]
The average density of particles is fixed by the thermodynamic relation
\begin{equation}
c=z\partial_{z}f(z),\label{eq:density-fugacity}
\end{equation}
which, written in terms of $z_{c}$ becomes
\begin{equation}
c=\frac{(1-\nu)z_{c}}{1-\nu\left(2-z_{c}\right)z_{c}}.\label{eq:density}
\end{equation}
 The density takes values in the range
\begin{equation}
c\in[0,1]\label{eq: density range}
\end{equation}
 as $z_{c}$ varies in the same range.

{\subsection{Correlation functions and stationary observables}}
To evaluate $s-$point correlation functions of the form $\left\langle \tau_{k_{1}}\dots\tau_{k_{s}}\right\rangle _{L}$,
where $\left\langle a\right\rangle _{L}$ is the notation for expectation
value of the random variable $a$ in the system with $L$ sites, one
has to insert the matrix
\[
\widehat{\tau}=\left(\begin{array}{cc}
0 & 0\\
0 & 1
\end{array}\right)
\]
into the product of transfer matrices in the places corresponding
to sites $k_{1},\dots,k_{s}$:
\begin{equation}
\left\langle \tau_{k_{1}}\dots\tau_{k_{s}}\right\rangle _{L}=\frac{\mbox{Tr}\left[T^{k_{1}}\widehat{\tau}T^{k_{2}}\widehat{\tau}\dots\widehat{\tau}T^{L-(k_{1}+\dots+k_{s})}\right]}{\mathcal{Z}_{L}(z)}.\label{eq: correlation}
\end{equation}
To evaluate expressions of this kind we also need the eigenvectors
of $T$,
\begin{eqnarray}
\mathbf{\!\!\!\!v}_{1}\!\! & =\!\! & \left[\begin{array}{c}
\sqrt{\frac{\left(1-z_{c}\right)\left(1-\nu z_{c}\right)}{1-\nu\left(2-z_{c}\right)z_{c}}}\\
\\
\sqrt{\frac{z_{c}\left(1-\nu\right)}{1-\nu\left(2-z_{c}\right)z_{c}}}
\end{array}\right]\!,\mathbf{v}_{2}\!=\!\left[\begin{array}{c}
-\sqrt{\frac{z_{c}\left(1-\nu\right)}{1-\nu\left(2-z_{c}\right)z_{c}}}\\
\\
\sqrt{\frac{\left(1-z_{c}\right)\left(1-\nu z_{c}\right)}{1-\nu\left(2-z_{c}\right)z_{c}}}
\end{array}\right]\!,\,\,\,\,\label{eq:eigenvectors}
\end{eqnarray}
corresponding to $\lambda_{1}$ and $\lambda_{2}$ respectively, which
are normalized to $\left\Vert v_{1}\right\Vert =\left\Vert v_{2}\right\Vert =1.$
In particular, one can verify the fact that the one-point correlation
function indeed yields the density (\ref{eq:density}) in the thermodynamic
limit
\begin{equation}
\left\langle \tau\right\rangle _{L}=\frac{\mbox{Tr}(\widehat{\tau}T^{L})}{\mathcal{Z}_{L}(z)}\to_{L\to\infty}(\mathbf{v}_{1},\widehat{\tau}\mathbf{v}_{1})=c.\label{eq: correlation one-point}
\end{equation}

Also, one can evaluate the average current as the probability of event
$A$, when a particle crosses the bond connecting site $L$ with site
$1$ at a time step. The particle can jump into the site that either
was empty after the previous step (event $B_{0}$) or has emptied
at the same step as a result of the shift of a cluster of size $k=1,2,\dots$
(event $B_{k}$). Thus, the current $j_{\infty}(c)$ maintained in
the infinite system at the particle density $c$ related by (\ref{eq:density})
with $z_{c}$ can be evaluated as
\[
j(c)=\mathbb{P}_{\infty}(A)=\sum_{k\geq0}\mathbb{P_{\infty}}(A\bigcap B_{k}),
\]
where the subscript in $\mathbb{P}_{\infty}$ refers to the infinite
system size. The finite size $L$ lattice version of the probabilities
under the sum are evaluated to
\begin{align}
 & \mathbb{P}_{L}(A\bigcap B_{k})=p\mu^{k}\langle\tau_{L}\tau_{1}\cdots\tau_{k}(1-\tau_{k+1})\rangle_{L}\label{eq:eq: P_L(A and B_k)}\\
 & =p\mu^{k}(\langle\tau_{L}\tau_{1}\cdots\tau_{k}\rangle_{L}-\langle\tau_{L}\tau_{1}\cdots\tau_{k+1}\rangle_{L})\nonumber \\
 & =p\mu^{k}z^{k}\left(\langle\tau_{1}\rangle_{L-k}\frac{\mathcal{Z}_{L-k}(z)}{\mathcal{Z}_{L}(z)}-z\langle\tau_{1}\rangle_{L-k-1}\frac{\mathcal{Z}_{L-k-1}(z)}{\mathcal{Z}_{L}(z)}\right)\nonumber
\end{align}
where $\mathbb{P}_{L}$ and $\langle\cdot\rangle_{L}$ are the notations
for the probability and expectation in the system of size $L$ and
we used the relation
\[
\langle\cdots\tau_{i}\tau_{i+1}\cdots\rangle_{L}\mathcal{Z}_{L}(z)=\langle\cdots\tau_{i}\cdots\rangle_{L-1}z\mathcal{Z}_{L-1}(z),
\]
that reduces the average of a function of configuration, vanishing
unless two given neighboring sites belong to the same particle cluster,
to a similar average of in the system with one site excluded.

\begin{figure}
\includegraphics[width=0.9\columnwidth]{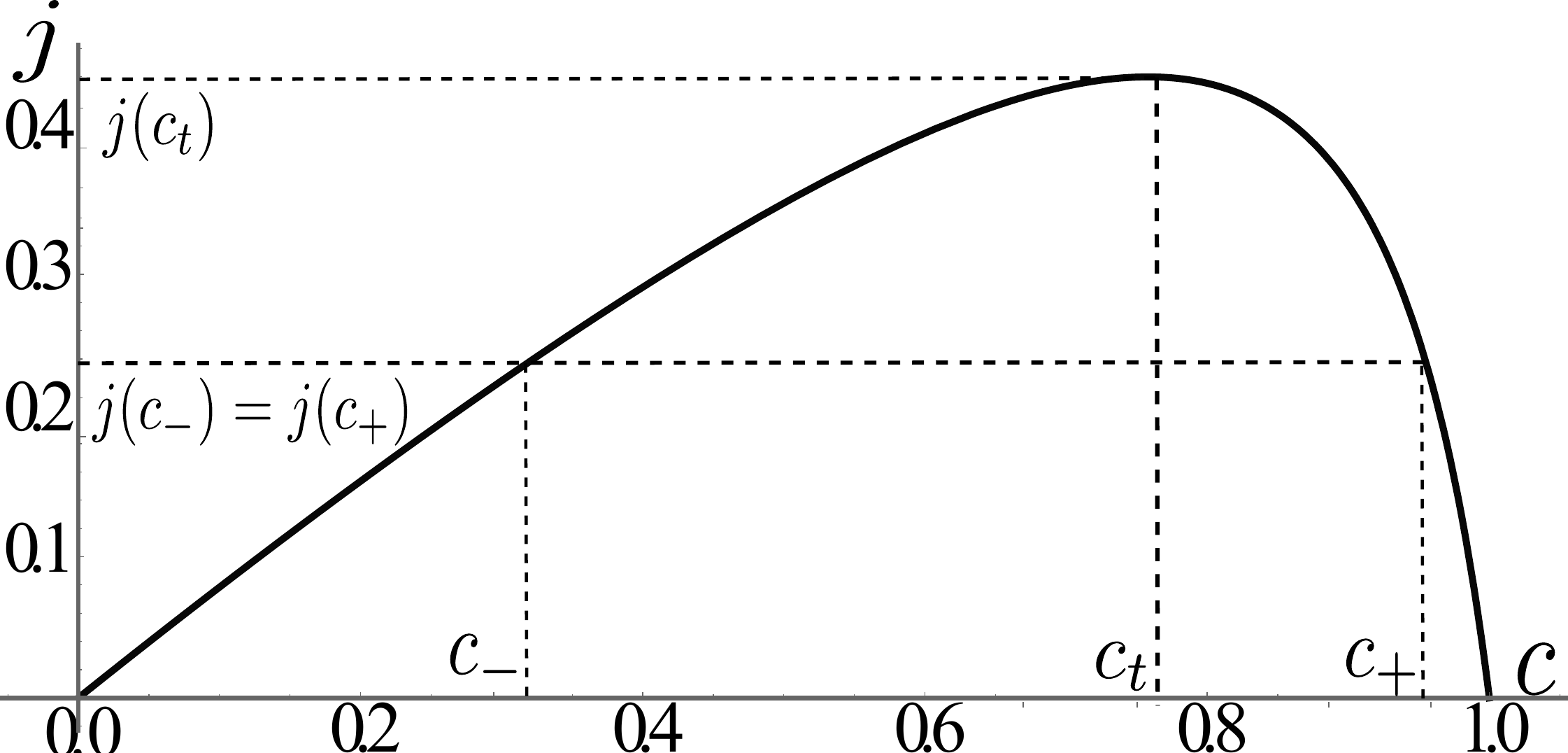}\caption{An example of current density plot for $p=0.8,\mu=0.9.$ The points
$c_{+},c_{-}$ and $c_{t}$ show a pair of left and right reservoir
densities corresponding to a point at the phase coexistence line and
the density of MC phase respectively. \label{fig:current density plot} }
\end{figure}

The $L\to\infty$ limit yields
\begin{align}
\mathbb{P}_{\infty}(A\bigcap B_{k}) & =p\mu^{k}\left(\frac{z}{\lambda_{1}}\right)^{k}c\left(1-\frac{z}{\lambda_{1}}\right)\label{eq:eq: P(A and B_k)}\\
 & =p\mu^{k}cz_{c}^{k}(1-z_{c}),\quad k=0,1,\cdots.\nonumber
\end{align}
Then, for the current we have
\begin{align}
j(c) & =pc(1-z_{c})\sum_{k\geq0}(\mu z_{c})^{k}\label{eq:j}\\
 & =\frac{\left(1-z_{c}\right)z_{c}(\mu-\nu)}{\left(1-\mu z_{c}\right)\left(\nu z_{c}^{2}-2\nu z_{c}+1\right)},
\end{align}
which is the expression previously obtained in \citep{woelki2005steady,derbyshev2015emergence}.
For all $0\leq\mu\leq 1$ and $0<p\leq1$ it is a concave function of the
density taking values $j(0)=j(1)=0$ at $c=0$ and $c=1$ respectively,
and hence having a single maximum in (\ref{eq: density range}), see
fig. \ref{fig:current density plot}. \\

{\subsection{ Injection and ejection probabilities}}
Let us evaluate the values of probabilities for a particle to enter
and exit the system in gTASEP on a segment with open ends, implying
that they are the same as the probability of similar moves in the
infinite system in translation invariant steady state at fugacity
$z_{c}$ (and hence at density (\ref{eq:density})). For particles
entering the system we can define the family of probabilities
\[
\alpha_{k}=\mathbb{P_{\infty}}(A|B_{k})=\frac{\mathbb{P_{\infty}}(A\bigcap B_{k})}{\mathbb{P}_{\infty}(B_{k})},\quad k=0,1,\dots
\]
where $\alpha_{0}$ is the probability for a particle to enter 
a site conditioned on the fact that this site remained empty after
the previous step, while $\alpha_{k}$ with $k>0$ is a similar probability
conditioned on the fact that the site has emptied after the one-step
shift of a cluster of $k>0$ particles at the same time step. While
the numerators were obtained in (\ref{eq:eq: P_L(A and B_k)},\ref{eq:eq: P(A and B_k)})
the denominators can be evaluated similarly
\begin{align*}
\mathbb{P}_{L}(B_{0}) & =\langle(1-\tau_{1})\rangle_{L},\\
\mathbb{P}_{L}(B_{k}) & =p\mu^{k-1}\langle\tau_{1}\cdots\tau_{k}(1-\tau_{k+1})\rangle_{L},\quad k\geq1.
\end{align*}
In the $L\to\infty$ limit this yields
\begin{align*}
\mathbb{P}_{\infty}(B_{0}) & =1-c\\
\mathbb{P}_{\infty}(B_{k}) & =p\mu^{k-1}cz_{c}^{k-1}(1-z_{c}).
\end{align*}
Thus we have
\begin{equation}
\alpha_{0}=\frac{pc(1-z_{c})}{1-c}=\frac{(\mu-\nu)z_{c}}{(1-\nu z_{c})}\label{eq: alpha_0}
\end{equation}
and
\begin{equation}
\alpha_{k}=\mu z_{c},\quad k\geq1.\label{eq: alpha_k}
\end{equation}
One can see that $\alpha_{k}$ does not depend on $k$, when $k\geq1$.
Thus, similarly to the bulk dynamics it is enough to define two entrance
probabilities
\begin{align}
\alpha & =\frac{(\mu-\nu)z_{c}}{(1-\nu z_{c})},\label{eq:alpha}\\
\tilde{\alpha} & =\mu z_{c},\label{eq:tilde alpha}
\end{align}
for the particle to enter the first site that was empty after the
previous step and that has emptied at the current step respectively.
They are parametrically related via the mutual dependence on the parameter
$z_{c}$. When $z_{c}$ varies in the range (\ref{eq:z_c range})
that corresponds to the densities in (\ref{eq: density range}), $\alpha$
and $\tilde{\alpha}$ take values in
\begin{align}
\alpha & \in[0,p],\quad\tilde{\alpha}\in[0,\mu],\label{eq:alpha range}
\end{align}
being monotonous functions of $z_{c}.$ Eliminating $z_{c}$ between
(\ref{eq:alpha}) and (\ref{eq:tilde alpha}), we obtain (\ref{eq:tilde alpha-1}).

Similarly, to evaluate the analogue of probability $\beta$ for the
particle to leave the system from the last site we should find the
probability for the particle to jump out of the site conditioned on
the presence of a particle at this site, which reads
\begin{align}
\beta & =\mathbb{P}(A|\{\tau_{L}=1\})\nonumber \\
 & =\frac{j(c)}{c}=\frac{(1-z_{c})(\mu-\nu)}{(1-\nu)(1-\mu z_{c})}.\label{eq:beta}
\end{align}
It also takes values in the range
\begin{equation}
\beta\in[0,p]\label{eq:beta range}
\end{equation}
for $z_{c}$ in (\ref{eq:z_c range}). Both $\alpha$ and $\beta$
are monotonous functions of $z_{c}$ (increasing and decreasing respectively)
and hence of the density.\\

\section{Phase diagram of  gTASEP\label{sec: Phase diagram}}

Let us use relations (\ref{eq:density},\ref{eq:j},\ref{eq:alpha}-\ref{eq:beta})
to establish the form of the phase diagram of gTASEP on a segment.
The mechanism of the stationary state selection in the driven diffusive
systems with the current with a unique maximum in the current-density
plot was described in \citep{krug1991boundary,kolomeisky1998phase}
using purely hydrodynamic arguments. Below we apply their findings
to construct the phase diagram of gTASEP.

{\subsection{Stationary state trivialization curve}}

First we note that when the values of $\alpha$ and $\beta$ satisfy
(\ref{eq:alpha range},\ref{eq:beta range}) the densities $c_{-}$and
$c_{+}$ of the left and right reservoirs associated with our BC are
obtained by eliminating $z_{c}$ between the $\alpha$ and $c$, eqs.(\ref{eq:density},\ref{eq:alpha}),
and $\beta$ and and $c$, eqs. (\ref{eq:density},\ref{eq:beta}),
which yields
\begin{align}
c_{-} & =\frac{\alpha^{2}(\mu-p)+\alpha p(1-\mu)}{\alpha^{2}(\mu-p)+p^{2}(1-\mu)}\label{eq:density alpha}
\end{align}
and
\begin{equation}
c_{+}=\frac{(p-\beta)(p-\beta\mu)}{p\left(p-\beta^{2}\right)+\beta\mu(\beta+(\beta-2)p)},\label{eq: density beta}
\end{equation}
respectively. A special curve in the phase diagram, at which the densities
of the left and right reservoirs are the same, $c_{+}=c_{-}$, can
be obtained by eliminating $z_{c}$ between (\ref{eq:beta}) and (\ref{eq:alpha}),
which yields
\begin{equation}
(1-\beta)(1-\alpha)=1-p.\label{eq: alpha-beta line}
\end{equation}
 At this curve the system on the segment looks like a part of the
infinite system in the translation invariant steady state. In this
case, the exact stationary measure similar to that of 1D Ising model
can be obtained as $L$-site correlation function at the infinite
lattice and expressed in terms of a product of $2\times2$ matrices,
sandwiched between two vectors $\boldsymbol{v}_{1}$ from (\ref{eq:eigenvectors}).
In particular, the macroscopic density profile is perfectly flat with
$c=c_{+}=c_{-}$ and the current is $j(c)$. The relation (\ref{eq: alpha-beta line})
was already obtained before for the cases of parallel and backward
sequential updates, where it defines the regime in which the representation
of the algebra used to construct the matrix product stationary state
trivializes. Remarkably, this relation is preserved at the arbitrary
values of $\mu$.

{\subsection{Triple point}}

As we move along the curve (\ref{eq: alpha-beta line}) from $(\alpha,\beta)=(0,p)$
to $(\alpha,\beta)=(p,0)$ the density $c=c_{-}=c_{+}$ continuously
grows from $c=0$ to $c=1$ as well as the associated fugacity $z_{c}$.
At the same time, the particle current $j(c)$ grows from $j(0)=0$
to its unique maximum $j_{\max}=j(c_{t})$ attained at the density
$c=c_{t}$, the unique root of the equation
\[
\frac{dj(c)}{dc}=0,
\]
in the density range (\ref{eq: density range}),
and then decreases back to $j(1)=0$. The point corresponding to $c=c_{t}$
is a triple point. Since the current attains its maximal value at
$c_{t}$ and this value is the largest that the bulk can sustain,
a further increase neither of $\alpha$ nor of $\beta$ causes an
increase of the current. Thus, the rectangle $(\alpha,\beta)\in[\alpha_{t},1]\times[\beta_{t},1])$
is the maximal current phase.   Remarkably, as follows from (\ref{eq: alpha-beta line})  its area does not depend on the value of $\mu$.

\begin{figure}
\noindent \centering{}\includegraphics[width=\columnwidth]{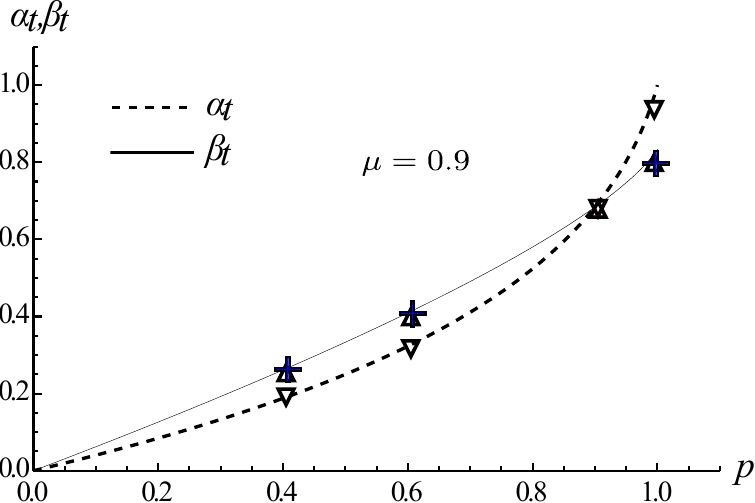} \caption{The values of $\alpha_{t}$ and $\beta_{t}$ at the triple point as functions of $p$ corresponding to $\mu=0.9$. The graphs cross at
	two points corresponding to PU and BSU cases. The  down-  and up-triangles, $\triangledown,\triangle,$ are their values measured in simulations. The  cross symbols, ${+}$, are the values of $\beta_t$ in gTASEP with left BC (\ref{eq: Standard BC}), which coincide with ones with Liggett's  BC within the numerical precision. \label{fig:alpha-beta-p}}
\end{figure}

\begin{figure}
\noindent \centering{}\includegraphics[width=\columnwidth]{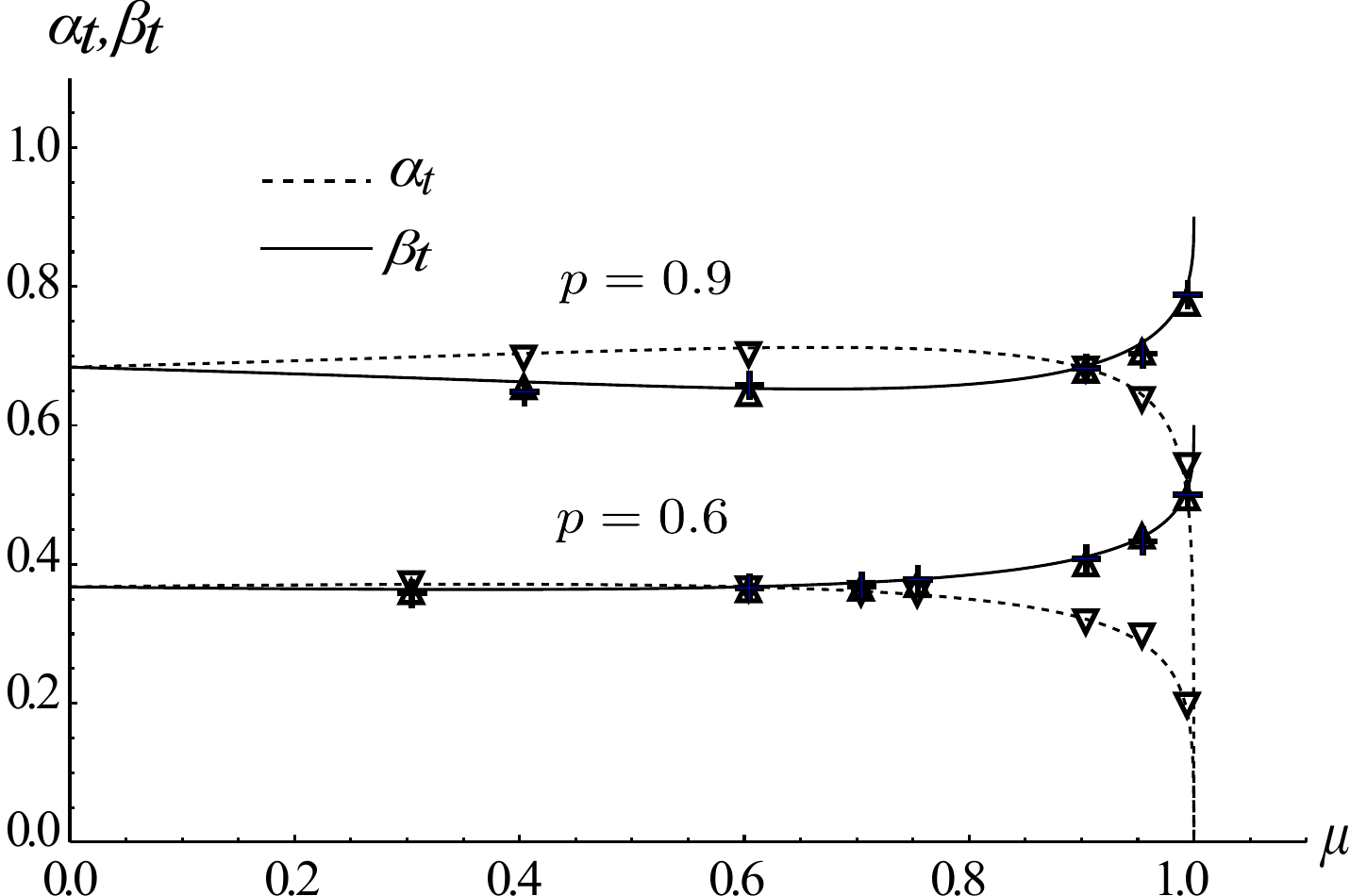}
\caption{The values of $\alpha_{t}$ and $\beta_{t}$ at the triple point as
functions of $\mu$ corresponding to $p=0.6$ and $p=0.9$. 
The graphs cross at
two points corresponding to to PU and BSU
cases. The  down- and up-triangles, $ \triangledown, \triangle,$ are their values measured in simulations. The  cross symbols, $+$, are the values of $\beta_t$ in gTASEP with left BC (\ref{eq: Standard BC}), which coincide with ones with Liggett's  BC within the numerical precision. 
 \label{fig:alpha-beta-mu}}
\end{figure}

The value of $c_{t}$ as well as $j_{\max}$,$\alpha_{t},\tilde{\alpha}_{t}$
and $\beta_{t}$ can be obtained by evaluating formulas (\ref{eq:density},\ref{eq:j},\ref{eq:alpha},\ref{eq:tilde alpha},\ref{eq:beta})
respectively at $z_{c}$ given by a unique root of the polynomial
\begin{equation}
P_\mathrm{T}(z)=\mu\nu(z^{4}-2z^{3}+2z^{2})-(\mu+\nu)z^{2}+2z-1\label{eq:trple point polynomial}
\end{equation}
in the range (\ref{eq:z_c range}), see Figs. \ref{fig:alpha-beta-p},\ref{fig:alpha-beta-mu}.
In two cases $\mu=0$ and $\nu=0$ ($\mu=p$) corresponding to PU
and BSU respectively $P_\mathrm{T}(z)$ simplifies to the quadratic polynomial,
and the explicit values of $c,j_{\infty}$ and $\alpha_{t},\tilde{\alpha}_{t}$
and $\beta_{t}$ at the triple point can easily be evaluated reproducing
the known expressions obtained before. In particular
\begin{equation}
\alpha_{t}=\beta_{t}=1-\sqrt{1-p}\label{eq:alpha_t, beta_t mu=00003D0,nu=00003D0}
\end{equation}
 in both cases. Remarkably, these are the only two cases, when $\alpha_{t}$
coincides with $\beta_{t}$. Furthermore, as follows from (\ref{eq:alpha_t, beta_t mu=00003D0,nu=00003D0}),
given $p$,  the values of $\alpha_{t}$ and $\beta_{t}$ are not
monotonous as functions of $\mu$ in the interval $\mu\in[0,p]$.
As $\mu$ increases, the value of $\alpha_{t}$ ($\beta_{t}$) starting
from (\ref{eq:alpha_t, beta_t mu=00003D0,nu=00003D0}) at $\mu=0$
first increases (decreases), but then turns back to (\ref{eq:alpha_t, beta_t mu=00003D0,nu=00003D0})
at $\mu=p$ and continues to decrease (increase) up to $\alpha_{t}=0$,
($\beta_{t}=p$) at $\mu=1$.

\begin{figure}
\noindent \begin{centering}
\includegraphics[width=0.9\columnwidth]{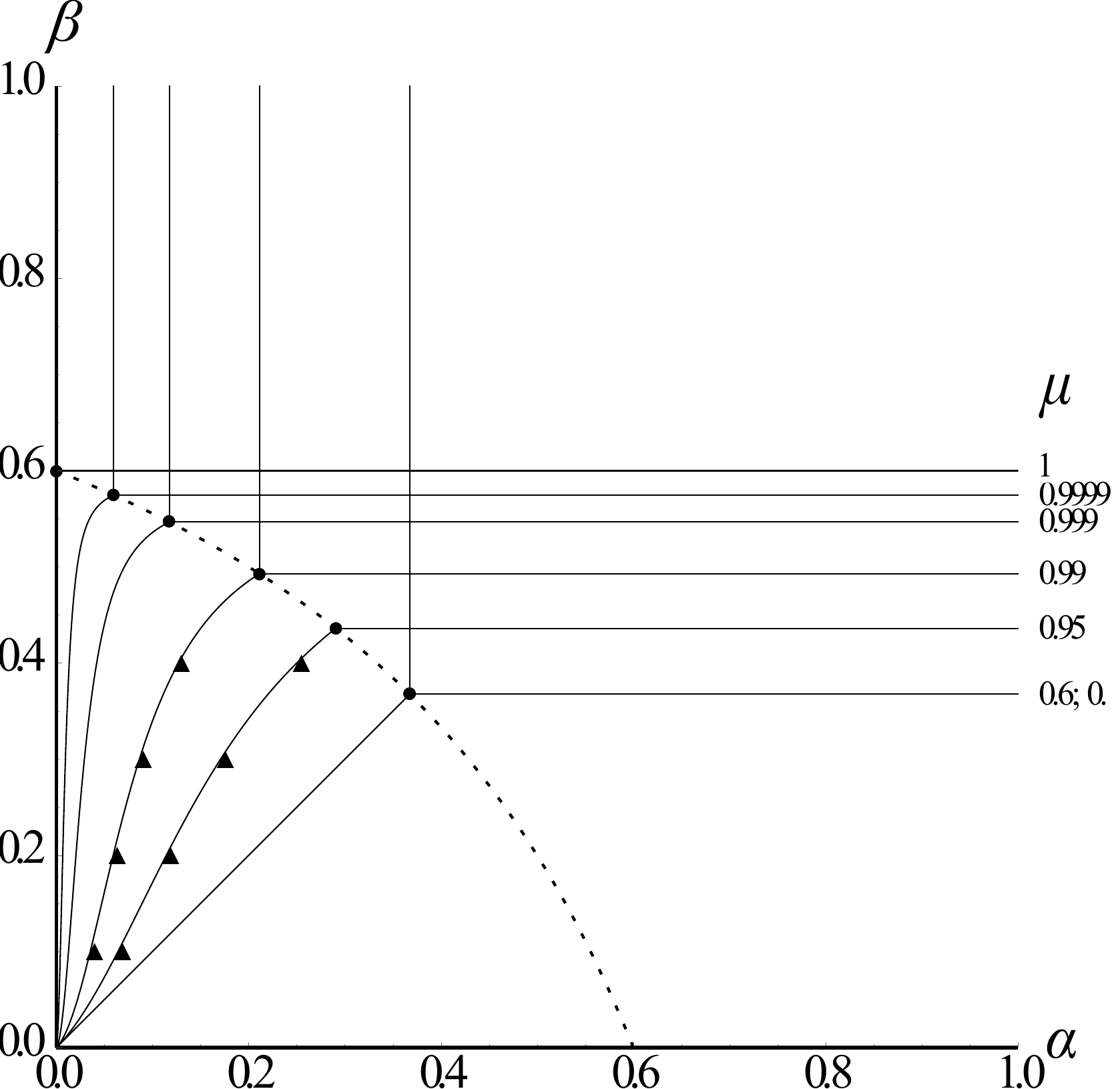}\\
\centerline{(a)}
\includegraphics[width=0.9\columnwidth]{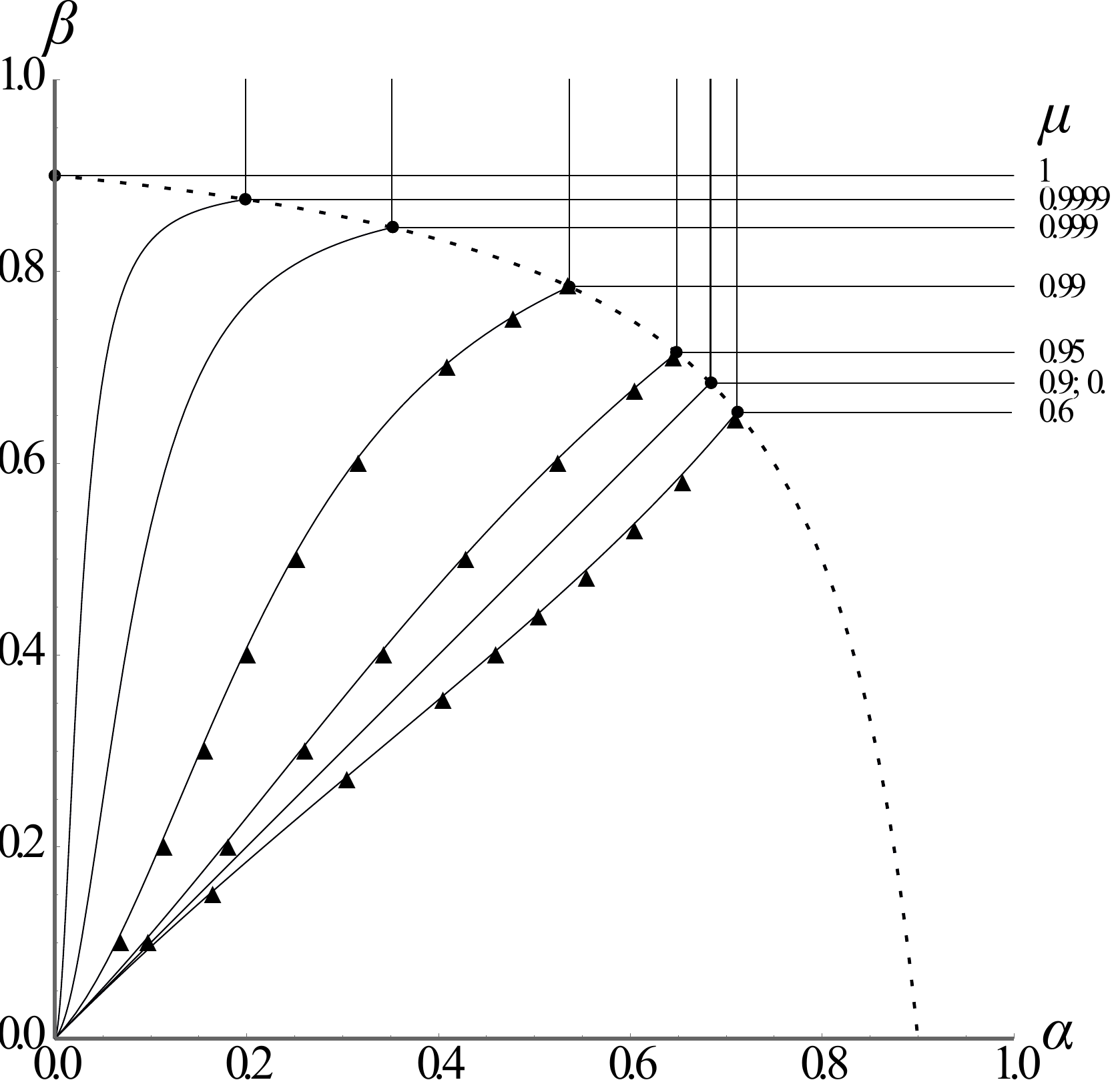}
\centerline{(b)}
\par\end{centering}
\noindent \centering{}\caption{The phase diagrams of gTASEP in $\alpha-\beta$ plane for (a) $p=0.6$ and (b) $p=0.9$
with various  values of $\mu$. The dashed curves are the lines
(\ref{eq: alpha-beta line}), where the stationary state trivializes.  The triangle symbols, $\blacktriangle$, are the  simulation results. 
\label{fig:The-phase-diagrams-gTASEP}}
\end{figure}

{\subsection{HD-LD phase coexistence line}}

Let us finally describe the HD and LD phases and, in particular, the
phase coexistence line between them. The LD (HD) phase is established,
when $j(c_{-})<\min(j(c_{+}),j_{\max})$ $(j(c_{+})<\min(j(c_{-}),j_{\max})$. It includes  
the points of the  phase diagram  under the curve (\ref{eq: alpha-beta line}), where the
boundaries are associated with reservoirs at densities $c_{-}$ and
$c_{+}$, as well as for larger values of $\beta$ ($\alpha$). The
particle density is equal to $c_{-}(c_{+})$ in almost all the system,
up to a close vicinity of the right (left) segment end, where the
density profile bends to reach the density of the left (right) reservoir.

The phase coexistence line connects the origin $(\alpha,\beta)=(0,0)$
to the triple point $(\alpha_{t},\beta_{t})$ and can be obtained
as a set of pairs $(\alpha,\beta)$ associated with densities $(c_{-},c_{+})$
solving equation
\begin{equation}
	j(c_{-})=j(c_{+}),
\end{equation}
see Fig. \ref{fig:The-phase-diagrams-gTASEP}.

In terms of fugacities $z_{-},z_{+}$ associated with densities $c_{-},c_{+}$
this equation is reduced to finding the 1D subset of $(z_{-},z_{+})\in[0,z_{t}]\times[z_{t},1]$,
where polynomial
\begin{align*}
P_{\mathrm{C}}(z_{-},z_{+}) & =\mu\nu z_{+}^{2}z_{-}^{2}-\mu\nu z_{+}z_{-}^{2}-\mu\nu z_{+}^{2}z_{-}\\
 & -z_{+}z_{-}(\mu+\nu-2\mu\nu)+z_{-}+z_{+}-1
\end{align*}
vanishes, which amounts to solving the quadratic equation, say for
$z_{+}$ in terms of $z_{-}$. Substituting solutions obtained for
$(z_{-},z_{+})$ into (\ref{eq:alpha},\ref{eq:beta}) respectively
we obtain the required parametric curve in $(\alpha,\beta)$ coordinates
as functions of $z_{-}$ varying in the range $0\leq z_{-}\leq z_{t}$.
The result simplifies significantly in the PU and BSU cases $\mu=0$
and $\nu=0$ respectively, yielding the known result, the straight
line segment $\alpha=\beta\in[0,1-\sqrt{1-p}].$

As one can see in fig. \ref{fig:The-phase-diagrams-gTASEP},
the shape of the phase coexistence line starting from the straight segment at $\mu=0$
 is bending as $\mu$  increases, first rotating clockwise, so that the area of LD phase  grows, 
 then rotating anticlockwise to return back to  the same straight segment at $\mu=p$. Then, it is again bending and moving
  towards the  $\beta$-axis, so that the low density phase disappears in the limit $\mu\to 1$.\\

{\subsection{Deterministic aggregation limit}}

The limit $\mu\to 1$  at fixed $0<p<1$ is associated with deterministic aggregation regime in \cite{derbyshev2015emergence}, in which  particles tend to  gather in large clusters. With Liggett-like BC we automatically have $\tilde{\alpha}\to 1$ in this limit
 for any $0<\alpha<1$, so that  in the stationary state all the lattice is occupied by a single cluster that sometimes (with probability $\beta$) shifts as a whole one step forward with immediate filling the emptied left-most site, so that the  density maintained is exactly $c=1$. To see the lower densities one should consider the joint limits $\mu\to 1$ and  $z_\pm \to 1 $   corresponding to $\alpha\to 0$ and $\beta\to p$ in LD and HD phases respectively.  
 
 Specifically, following  \cite{derbyshev2021nonstationary}  one can consider the family of scalings  parameterized  by an exponent $\mathfrak{b}\in (0,1)$\footnote{Notation $\mathfrak{b}$ is used in place  of $\beta$ of  \cite{derbyshev2021nonstationary}  since the latter is already reserved for the exit probability.} setting 
 \begin{equation}
 	z_c=1-\zeta (1-\mu)^{\mathfrak{b}},
 \end{equation}
where $\zeta>0$ is a scaling variable. Under this scaling we have
 \begin{eqnarray}
 c&=&\left\{
 \begin{matrix}
   (1-\mu)^{1-2\mathfrak{b}}  \left(\zeta^2 (1-p)\right)^{-1} & \mathfrak{b}<1/2  \\
 	 (1+\zeta ^2(1-p))^{-1} & \mathfrak{b}=1/2\\
	1-(1-p)\zeta^2(1-\mu)^{2\mathfrak{b}-1},& \mathfrak{b}>1/2
 	\end{matrix}
  \right.\label{eq: c mu->1}\\
   &&+\,O\left((1-\mu)^{2|2\mathfrak{b}-1|},(1-\mu)^{(|2\mathfrak{b}-1|+1)/2}\right),\notag
 \end{eqnarray}i.e.,  $\mathfrak{b}=1/2$  corresponds to  the density  ranging in $c\in (0,1)$ as $\zeta\in (0,\infty)$ with   values away from $c=0$ and $c=1$, while $\mathfrak{b}<1/2$  and $\mathfrak{b}>1/2$ are responsible  for the  close vicinities of the  extreme  density values respectively.  

Corresponding entrance and exit probabilities will than scale as 
\begin{eqnarray}
\alpha& =&\frac{p(1-\mu)^{1-\mathfrak{b}}}{\zeta(1-p)}+O\left((1-\mu),(1-\mu)^{2(1-\mathfrak{b})}\right), \label{eq: alpha mu->1}\\
\tilde{\alpha}&=&\left(1-\zeta (1-\mu)^{1-\mathfrak{b}}\right)+   \label{eq: tilde alpha mu->1} O\left((1-\mu)\right),\\
	\beta&=& p\left(1-\zeta^{-1}(1-\mu)^{1-\mathfrak{b}}\right)  \label{eq: beta mu->1} \\
	&&\,\,\,\,\,\,\,\,\,\,\,\,\,\,\,\,\,\,\,\,\,\,\,\,\,\,\,\,\,\,\,\,+\,O\left((1-\mu),(1-\mu)^{2(1-\mathfrak{b})}\right). \notag
\end{eqnarray}
in LD and HD phases respectively.  The formulas (\ref{eq: alpha mu->1},\ref{eq: tilde alpha mu->1}) and (\ref{eq: beta mu->1}) are the 
entrance and exit probabilities that maintain the densities (\ref{eq: c mu->1}) in LD and HD phases as $\mu\to 1$, provided that $\alpha$  and $\beta$ are  less than $\alpha_t$ and $\beta_t$ respectively. 

To characterize  the triple point   we study the roots of  $P_\mathrm{T}(z)$, (\ref{eq:trple point polynomial}), in the limit $\mu\to 1$. When $\mu=1$, this polynomial has a triple root $z=1$. Then, examining the expansion of these roots in  powers of $(1-\mu)$ we find the fugacity $z_c$  given by a single real root located in the range (\ref{eq:z_c range}) to be   
\begin{equation}
	z_c=1-\frac{(1-\mu)^{2/3}}{(2(1-p))^{1/3}}+O(1-\mu).
\end{equation}
Respectively, the triple point density and entrance and exit probabilities are given by formulas (\ref{eq: c mu->1}-\ref{eq: beta mu->1})  with $\mathfrak{b}=2/3$  and $\zeta=(2(1-p))^{-1/3}+O(1-\mu)^{1/3}.$

Finally, we note that  the crossover between  the KPZ and deterministic aggregation regimes studied in \cite{derbyshev2015emergence,derbyshev2021nonstationary}  corresponded to  such a  space scale,  
in which  one typically observes  finitely many huge particle clusters.   This fact suggested  a specific scaling of either the system size, when the stationary state on the ring was studied,    or spacial units, for  the non-stationary gTASEP in the  infinite system.     In our case, we also expect that the  change of the scaling behavior of  the particle current fluctuations will be observed in a joint limit $\mu \to 1$ and $L\to \infty$, such that the mean length per particle cluster  $c^{-1} \langle l_{\mathrm{cl}} \rangle$ is comparable   with the system size $L$.  Here  $$\left\langle l_{cl}\right\rangle=(1-z_c)^{-1}\simeq \sqrt{c/[(1-c)(1-\mu)]}$$ is the mean stationary cluster size  at the density $c$.
Hence, we conjecture  that the crossover will be governed by the scaling variable
\begin{equation}
	\tau = L (1-\mu)^{|\mathfrak{b}-\frac{1}{2}|+\frac{1}{2}}
\end{equation}
that stays finite in the limit under consideration, in the same way as it was in  \cite{derbyshev2015emergence,derbyshev2021nonstationary}, where the system crossed over from KPZ to fully correlated Gaussian 
motion of particles within a single giant cluster.\footnote{For details see the discussion of transitional scaling in \cite{derbyshev2021nonstationary}.}   In particular, we expect that in the vicinity of the triple point the current large deviation function  will  cross over from the form obtained in \cite{gorissen2012exact} to the pure Gaussian one as $\tau$ varies in the range $\tau\in[0,\infty)$.\\

\section{Simulation  results \label{sec: Simulation results}}

 To verify some of the analytical findings,  we also executed an independent numerical study of the steady state
properties of the gTASEP, equipped with Liggett's BC, defined by probabilities $\alpha,\beta$ and additional probability 
$\tilde{\alpha}$ given by formula
(\ref{eq:tilde alpha-1}). We performed numerical simulations of the system with the following parameters. A system of $L=200$
sites was studied. $N_{\tau}=800 000$ time steps were omitted to
ensure that the system has reached stationary state. The length of
the time sequences we used for recording the physical quantities
(e.g., density distribution along the system length $c(x)$,
$x\in (0,1]$ bulk density $c_b$, particle current $j$, etc.)
was $N_t=2^{21}$ , and averaging of over $N_{\text{ens}}=100$
independent simulation runs of the system was performed.

\begin{figure}
	\begin{center}
		\includegraphics[width=\columnwidth]{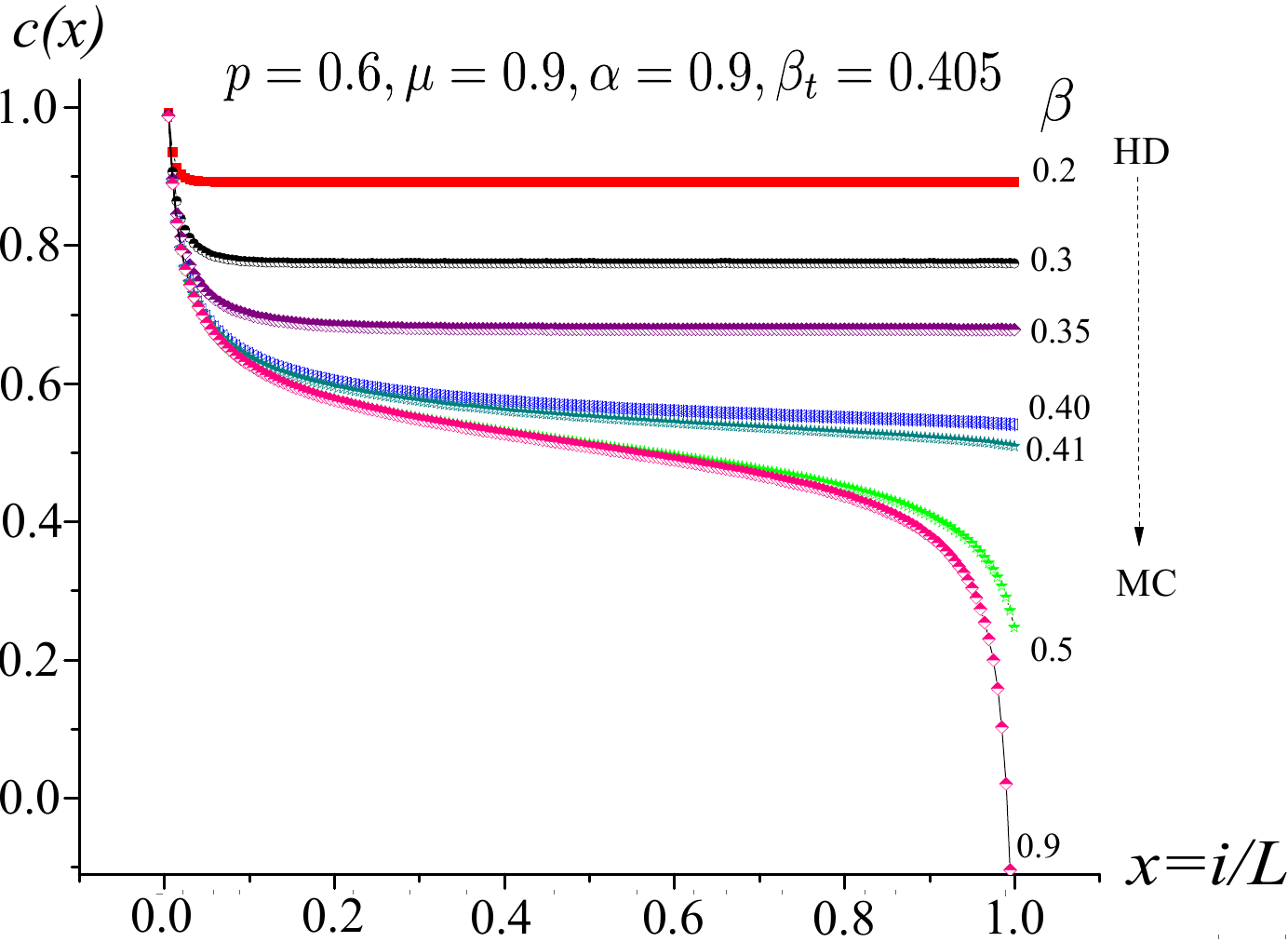}\\
		(a)
		\includegraphics[width=\columnwidth]{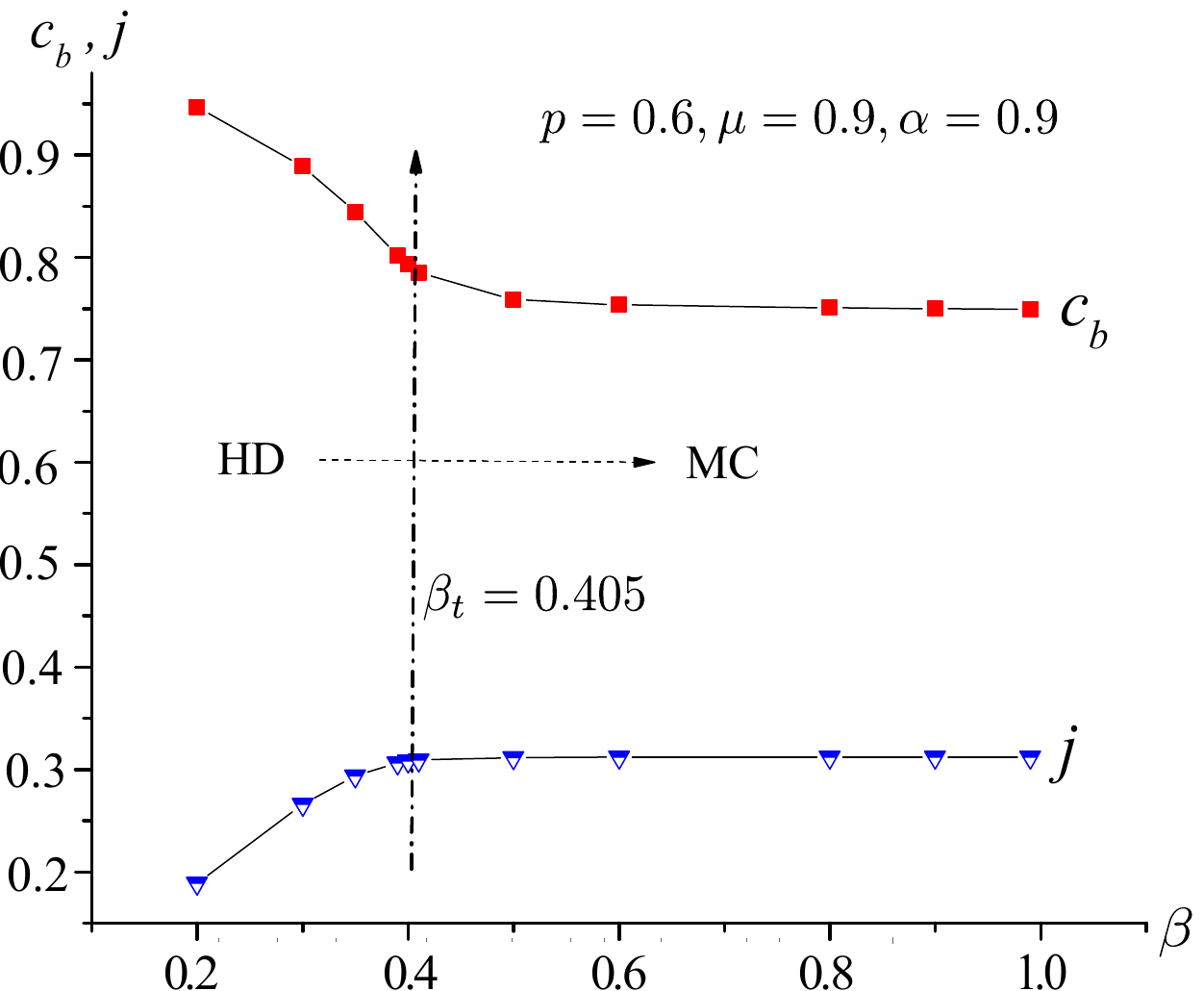}\\
		(b)
	\end{center}
	\caption{ \label{fig:HD-MC} Manifestation of HD-MC transitions. Numerical  results for	 different values of $\beta$ at fixed $p=0.6,\mu=0.9, \alpha =0.9>\alpha_t=0.3228$: 
		(a)   density profiles $c(x),x\in(0,1]$ along the system;
		(b) dependence of the current  $j$  and the bulk density $c_b$ on $\beta$ at fixed  $\alpha,p,\mu$.
		}
\end{figure}

\begin{figure}
\begin{center}
	\includegraphics[width=\columnwidth]{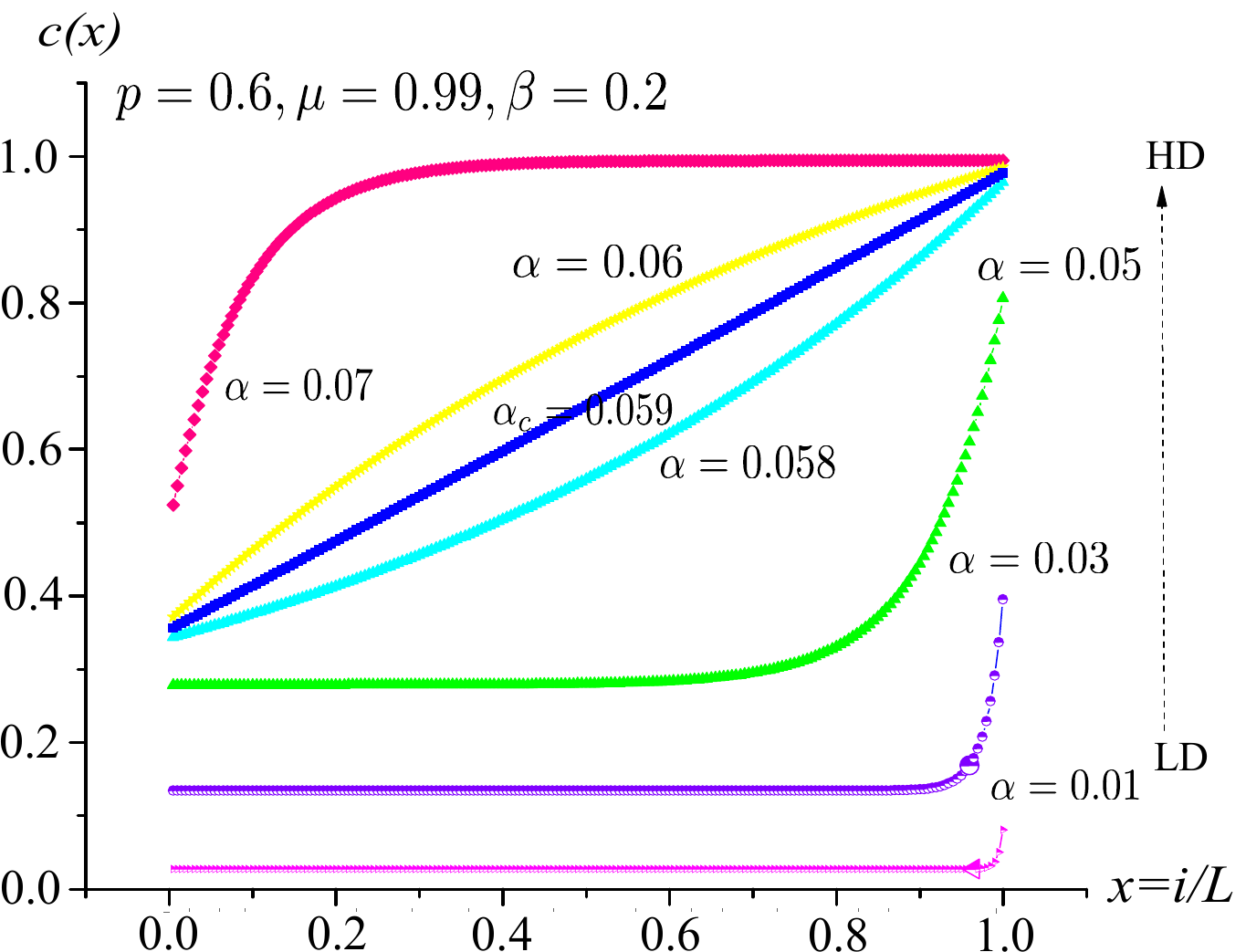}\\
	(a)
\includegraphics[width=\columnwidth]{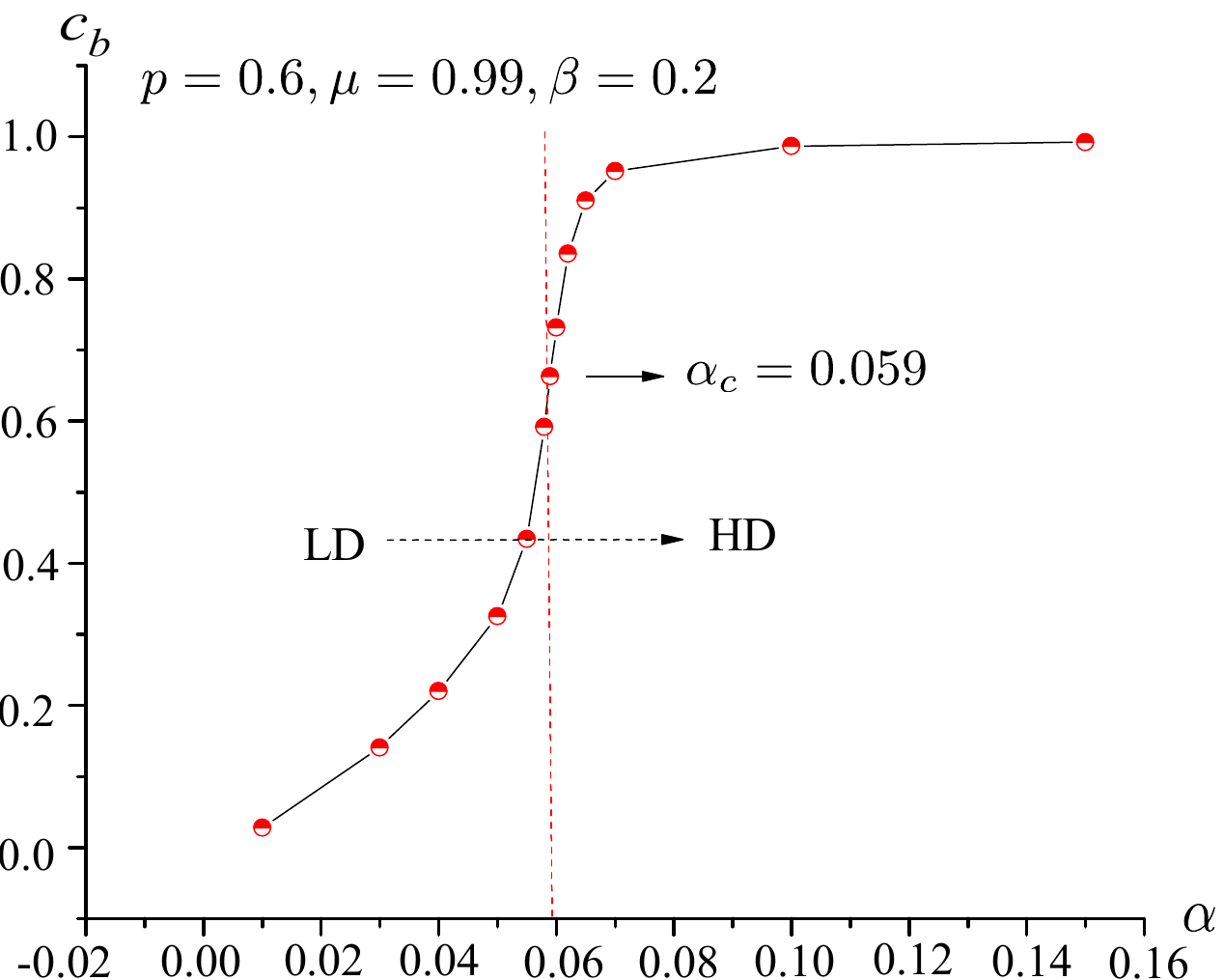}
(b)
\end{center}
\caption{Manifestation of  HD-LD  transitions. Numerical results for 	different values of $\alpha\in  (0.01,0.07)$  at fixed   $p=0.6, \mu=0.99 , \beta=0.2<\beta_t=0.49296$.   (a) Density profiles  for different values of $\alpha$. The profile at  the coexistence line is perfectly straight.  The change of the density profiles shape is used to determine the coexistence line coordinates. (b) Change of the bulk density $c_b$  at fixed $\beta$ as $\alpha$ crosses its value $\alpha_c$ at the HD-LD coexistence line. The transition is a bit ``smoothed'' since the system is finite.
\label{fig:LD-HD}}
\end{figure}

In Fig.~\ref{fig:HD-MC} we show  the density profiles $c(x),x\in [0,1]$,   obtained from the simulations in various regimes.   
In LD (HD) phase they are typically flat in the whole system except for a close vicinity of the right (left) end, where the profile bends to reach the density of the corresponding  reservoir. This confirms the fact that the stationary state of gTASEP in LD (HD) phase in a finite segment  is close to that  of the infinite system almost in the whole system.  In MC phase the profile maintains the MC density in the bulk and bends close to both ends of the segment to reach the densities of the reservoirs. As usual, the $L$-dependent size of the domain, where the density deviates from its bulk value, is much larger in MC phase than in HD and LD phases, resulting in the S-shaped profile in the finite system.  In Fig.~\ref{fig:HD-MC}(a)  the density profiles at fixed $p,\mu$  and $\alpha>\alpha_t$ are shown for seven values of $\beta$.  As one can see, for low
values of $\beta$ one has almost  constant  density profiles bending near the left end and after reaching the triple point $\beta_t$ the profiles gradually transform to the MC $S$-like shape. 

In order to determine the triple point values  $\alpha_t$ and $\beta_t$
for any given pair of values of the parameters $p,\mu$ we used
the dependence of the current $j$ on $\alpha$ (respectively, on
$\beta$) at fixed $\beta$ (respectively, fixed $\alpha$),
similarly to that in \cite{bunzarova2019one}. To illustrate that,
in Fig.  \ref{fig:HD-MC}(b) 
we present simulation results  for the dependence of the current $j$ and the bulk
density $c_b$ on $\beta$ at fixed $\alpha, p,\mu$. On the left side of the vertical dashed line
the system is in HD phase (characterized by high values of
$c_b$ and low values of $j$). After the transition point at
$\beta_t$  (on the right side of the vertical dashed line) the
dependence of the current $j$ on $\beta$ (and $c_b$) reaches a
plateau, signaling that the system is in the MC phase. 

We obtained the values of $\alpha_t$   and  $\beta_t$  at
fixed $p=0.9$  and $p=0.6$  for different values of $\mu$, Fig.  \ref{fig:alpha-beta-mu}, 
and at fixed $\mu=0.9$  for several values of
$p$, Fig. \ref{fig:alpha-beta-p}. As can be seen from figures, the numerical  values of
$\alpha_t$  and $\beta_t$  determined with precision $\pm 0.005$ corroborate with high
accuracy  the obtained analytical results.

If we fix the  values of $\alpha_t$ ($\beta_t$) below the one at the triple point and vary $\beta_t$ ($\alpha_t$) in its range, we will observe   the bulk density jump that corresponds to the LD-HD first order phase transition.    In Fig.~\ref{fig:LD-HD}(a) we show the profiles  at fixed $p,\mu$  and $\beta<\beta_t$  at seven different values. One can see that almost constant density profiles bending at the right end at small $\alpha$ transforms to those bending at the left end at large $\alpha$  through the point, where the density profile  is an exact straight line connecting the densities of the reservoirs. This is  the point at the coexistence line. We used this fact to locate  points of the coexistence line in simulations data. Also the bulk density $c_b$ that is expected  to make a sharp jump from LD to HD density value at this point, see Fig.  \ref{fig:LD-HD}(b). One can see in Fig. \ref{fig:The-phase-diagrams-gTASEP} that the  points at the  coexistence line obtained from the simulations are in excellent agreement with the analytically predicted curves.

\begin{figure}
	\vspace{0.5cm}
	\includegraphics[width=0.3\columnwidth]{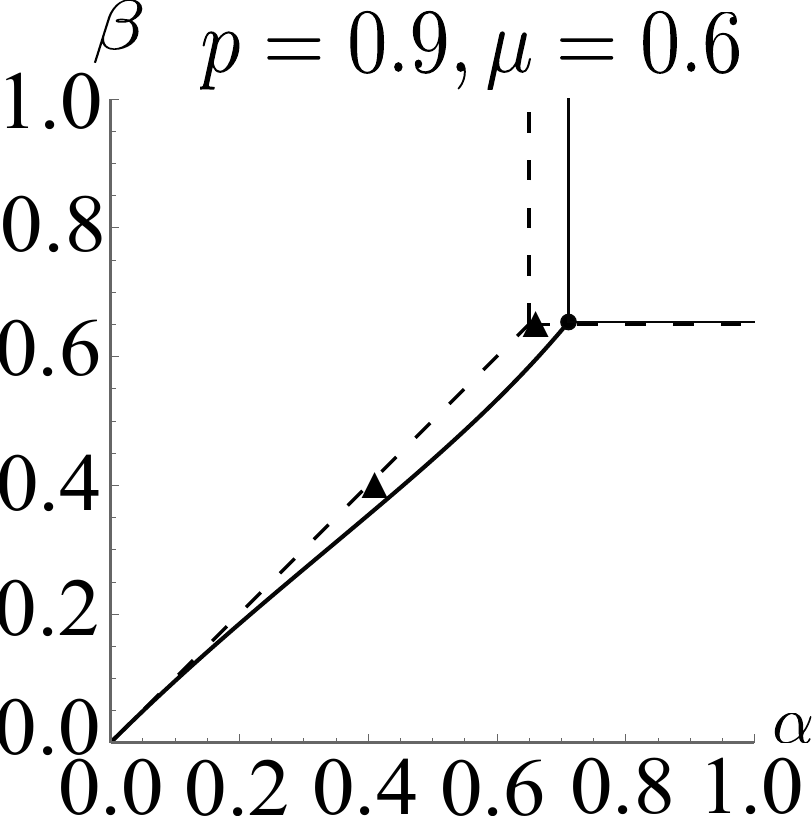}\,\,
	\includegraphics[width=0.3\columnwidth]{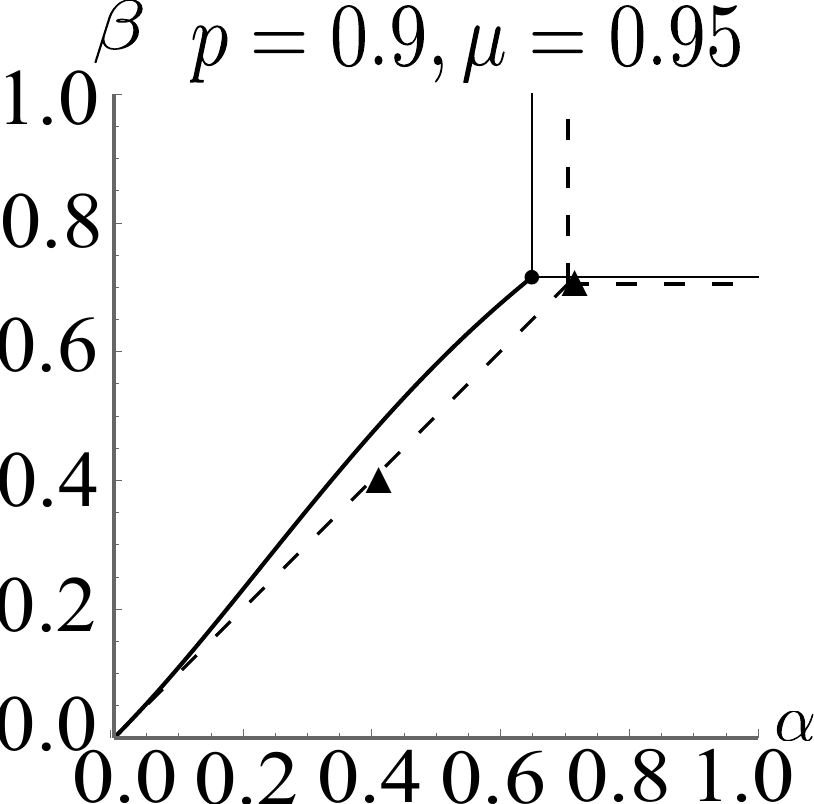}\,\,
	\includegraphics[width=0.3\columnwidth]{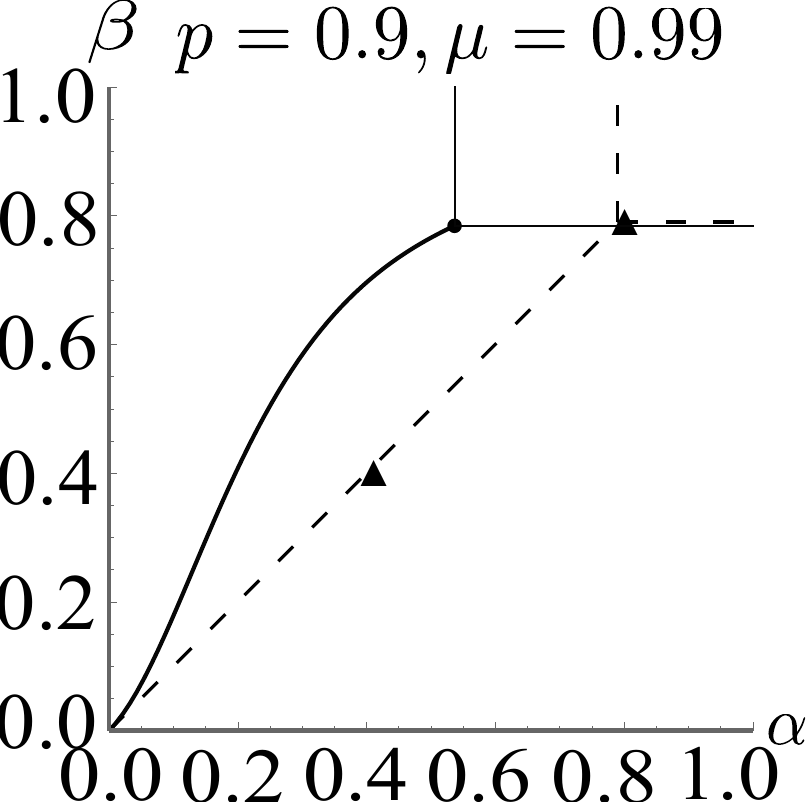}
	\caption{ \label{fig:v1-v2}Comparison of the phase diagrams of gTASEP with Liggett-like BC (solid  lines) and those with BC (\ref{eq: Standard BC}) (dashed lines) at three different values of $p,\mu$. Black triangles $\blacktriangle$  are the simulations results. We show the triple point and yet another point illustrating that the HD-LD phase coexistence line of  in gTASEP with BC (\ref{eq: Standard BC}) is straight. One can see that the triple point of gTASEP with BC  (\ref{eq: Standard BC}) is to the left from that   with Liggett-like BC   when $p>\mu$ and is to the right, when $\mu>p$.}
\end{figure}

Finally,  it is interesting to compare the phase diagrams obtained for gTASEP with Liggett-like BC with those for the previously studied  model with  with BC (\ref{eq: Standard BC}), see Fig. \ref{fig:v1-v2}. First, we note that the unlike the phase diagram of gTASEP with Liggett-like BC, the LD-HD phase coexistence line obtained with BC (\ref{eq: Standard BC}) is a straight line. Since we do not have an analytic theory of gTASEP with BC (\ref{eq: Standard BC}) we do not know the reason for this behavior. 

Also we note  that since the HD and MC  phases are governed by the right reservoir and the bulk  respectively the value of $\beta_t$ does not depend on the rules of update of the left end and thus is the same for the two types of BC. Indeed, as  can be seen in Figs. \ref{fig:alpha-beta-p},\ref{fig:alpha-beta-mu}, the  values of $\beta_t$ measure with the  two different  BC coincide within the numerical precision.  On the other hand, as explained in the end of section \ref{sec: Cluster dynamics},   BC (\ref{eq: Standard BC})  
enhance   the effective flow into the system  when  $\alpha<p$ and suppress it  when $\alpha>p$  in the repulsive regime, $p>\mu$,  and vice versa in the attractive regime,   $p<\mu$.  Therefore the value of $\alpha_t$ is greater (less) for BC (\ref{eq: Standard BC}) than for Liggett-like BC, when $p>\mu$ ($\mu>p$). This in turn means that the area of MC phase is  greater for  BC (\ref{eq: Standard BC}) than for Liggett-like BC in the repulsive regime and vice versa in the attractive one.

\section{Conclusion and discussion \label{sec: Conclusion and discussion}}
To conclude, we  proposed new boundary conditions for the TASEP with generalized update on an open segment. They mimic the infinite reservoirs  in translation invariant steady states attached to the segment ends. With these boundary conditions the phase diagram can be constructed explicitly. We confirmed our analytic predictions by extensive numerical simulations and compared them with the previous results on the same model with  boundary conditions defined in different way.   Of course, the most interesting part of the research subject is far  beyond the phase diagram. In particular, this would be the  fluctuation statistics of particle density and current  and especially the universal limit laws describing the large scale behavior of the system.  For this, we need an exact solution  at least for the steady state, for which the matrix product is known to be the main tool. At the moment the trivial $2\times 2$ matrix product stationary state exists only at the special line of the phase diagram, where the segment looks as a part of the infinite translation invariant stationary system. Also,  the matrix product solution is available for two particular limits of the generalized model we consider.  This makes us  expect that  a generalized matrix algebra that provides the matrix product stationary state of our system can also be obtained, and the proposed boundary conditions are the first candidate for obtaining its tractable representation. 
Then, our results  would serve as  the initial  test of the exact solution.  This is the next step for further investigation. We have also speculated about the crossover between the KPZ regime and  the deterministic aggregation limit and conjectured  the form of the scaling parameter controlling  the transition.     The scaling functions describing the crossover  could also be studied starting from  the exact solution. This would provide us with an  example of breaking up  the KPZ universality in the open system, when the interaction causes an unbounded growth of spatial correlation length.

\begin{acknowledgments}
AMP thanks Institute of Mechanics of the Bulgarian Academy of
Sciences for hospitality. The paper was supported  by the
Plenipotentiary Representative of the Bulgarian Government at the
Joint Institute for Nuclear Research, Dubna, theme 
01-3-1137-2019/2023. NCP and NZB thankfully acknowledge partial
financial support by the Bulgarian MES through the Bulgarian
National Roadmap on RIs (\rn{DO}1-325/01.12.2023).
\end{acknowledgments}

\bibliographystyle{apsrev4-2}
\bibliography{current_revtex}

\end{document}